\documentclass[aps,prx,amssymb,amsmath,superscriptaddress,twocolumn,10pt]{revtex4-1}
\usepackage[T1]{fontenc}
\usepackage[latin9]{inputenc}
\usepackage{amsmath}
\usepackage{graphicx}
\usepackage[monochrome]{color} 
\usepackage{longtable}
\usepackage{subfigure}
\usepackage{soul}

\def\ket#1{|{#1}\rangle}

\def\e{\mathrm{e}}
\def\ii{\mathrm{i}}
\def\d{\mathrm{d}}

\newcommand{\lyxdot}{.}

\begin{document}

\title{Phase Transitions in $Z_n$ Gauge Models: Towards Quantum Simulations of the Schwinger-Weyl QED}

\author{Elisa Ercolessi}
\affiliation{Dipartimento di Fisica e Astronomia dell'Universit\`a di Bologna, I-40127 Bologna, Italy}
\affiliation{INFN, Sezione di Bologna, I-40127 Bologna, Italy}

\author{Paolo Facchi}
\affiliation{Dipartimento di Fisica and MECENAS, Universit\`{a} di Bari, I-70126 Bari, Italy}
\affiliation{INFN, Sezione di Bari, I-70126 Bari, Italy}

\author{Giuseppe Magnifico}
\affiliation{Dipartimento di Fisica e Astronomia dell'Universit\`a di Bologna, I-40127 Bologna, Italy}
\affiliation{INFN, Sezione di Bologna, I-40127 Bologna, Italy}

\author{Saverio Pascazio}
\affiliation{Dipartimento di Fisica and MECENAS, Universit\`{a} di Bari, I-70126 Bari, Italy}
\affiliation{Istituto Nazionale di Ottica (INO-CNR), I-50125 Firenze, Italy}
\affiliation{INFN, Sezione di Bari, I-70126 Bari, Italy}

\author{Francesco V. Pepe}
\affiliation{INFN, Sezione di Bari, I-70126 Bari, Italy}

\begin{abstract}
We study the ground-state properties of a class of $\mathbb{Z}_n$ lattice gauge theories in $1+1$ dimensions, in which the gauge fields are coupled to spinless fermionic matter. These models, stemming from discrete representations of the Weyl commutator for the $\mathrm{U}(1)$ group, preserve the unitary character of the minimal coupling, and have therefore the property of formally approximating lattice quantum electrodynamics in one spatial dimension in the large-$n$ limit. The numerical study of such approximated theories is important to determine their effectiveness in reproducing the main features and phenomenology of the target theory, in view of implementations of cold-atom quantum simulators of QED. In this paper we study the cases $n=2\div 8$ by means of a DMRG code that exactly implements Gauss' law. We perform a careful scaling analysis, and show that, in absence of a background field, all $\mathbb{Z}_n$ models exhibit a phase transition which falls in the Ising universality class, with spontaneous symmetry breaking of the $CP$ symmetry. We then perform the large-$n$ limit and find that the asymptotic values of the critical parameters approach the ones obtained for the known phase transition the zero-charge sector of the massive Schwinger model, which occurs at negative mass.
\end{abstract}

\maketitle

\section{Introduction }

The advent of quantum technologies has opened unprecedented perspectives for the simulation of difficult problems, whose solution depends exponentially on input size. In particular, there is the intriguing possibility to explore lattice gauge theories~\cite{rothe1992lattice,montvay1997quantum}, such as quantum electrodynamics (QED) in 1D. This represents a complex and historical problem, that was investigated by  a number of outstanding physicists in the 70's~\cite{wilsonlgt,kogut1975hamiltonian,susskind1977lattice,kogut1979introduction}. 

The possibility of using quantum simulators to tackle these problems has been made possible by recent developments in low-temperature physics and atomic control techniques in optical lattices. Many proposals have been put forward in the literature to use ultra-cold atomic optical lattice systems to simulate many body models~\cite{bdz,qsim1,qsim2,qsim3,qsim4}, as well as Abelian and non-Abelian lattice gauge theories~\cite{simul1,simul4,simul3,simul5,simul6,simul7,simul9,simul11,zoharreview} and in particular quantum electrodynamics in 1D, that appears as a realizable option in the not-too-distant future~\cite{simul10,KCB,NEFMPP,simul12}. 

The key idea is that cold-atom quantum simulators make possible the implementation of matter fields in presence of artificially designed gauge fields by suitably identifying the gauge degrees of freedom with the internal (for example spin) states of the atom. The first experiments with fermions in presence of such ``synthetic''  fields have already been proposed and performed, offering very promising perspectives~\cite{cmr, ytterbium,mpi_sun,fallani,lcd}. Also, a first experiment reproducing 1D QED with few qubits has been reported~\cite{martinez2016}.

In general, the numerical study of approximated lattice-gauge theories is of paramount importance to determine their effectiveness in reproducing the main features and phenomenology of the target theory. Novel quantum-inspired numerical techniques, such as DMRG- and MPS-based algorithms~\cite{DMRG1,MPS}, fully exploit the entanglement of the states that contribute to the dynamics, and are able to reduce the computational cost by suitably tailoring the relevant (effective) subspaces in the Hilbert space.
From the theoretical point of view, an approach based on quantum simulators paves the way towards a number of problems that were traditionally difficult to analyze, such as the investigation of possible phase transitions, non-perturbative phenomena and dynamical aspects~\cite{simul2,simul6,BCJC,rico2016,montangero2015,schwinger_mps,buyens_prx,buyens}. \\
One of the main problems one has to face in encoding a lattice gauge-theory model in a cold-atom system is the fact that the number of states of the gauge field, sitting on the links between lattice sites, must be finite. In the literature, this has been accomplished in different ways. In~\cite{simul2,rico2016}, $U(1)$-gauge fields have been replaced by spin variables, which allow for finite dimensional (but non-unitary) representations of the canonical commutation relations. For one-dimensional QED this amounts to consider the so called quantum link model~\cite{qlm1,qlm2,qlm3,qlm4wiese}. Otherwise, one can try to discretize the gauge group, with the advantage of preserving the commutators, and guaranteeing at the same time the unitary character of the minimal coupling structure. This has been done for example in~\cite{KCB, buyens}, where the gauge group $U(1)$ was replaced by the discrete group $\mathbb Z$, which however admits only infinite dimensional representations. In such a case, the Hilbert space describing the gauge degrees of freedom must then be truncated to perform numerical simulations~\cite{buyens}.

In this paper we will follow a different strategy, that arises from the fact~\cite{NEFMPP} that the Weyl group commutation relations arising form the canonical commutators of the U(1) gauge fields admit discrete and finite dimensional representations, thus yielding a discrete and {\it finite} implementation of the gauge group commutator at the Hamiltonian level. This is important for two reasons. First, numerical simulations do not introduce any further approximation except for standard finite size effects, that can be dealt with by looking at scaling properties. Second, one can think of experimentally simulating {\color{red} the} class of models derived by such strategy, and described in the following in an exact way~\cite{NEFMPP}, by means of trapped cold atoms with synthetic degrees of freedom that satisfy periodic boundary conditions~\cite{aoc,cmr}.

We will consider $\mathbb{Z}_n$ Abelian gauge theories, with varying $n$, in which the gauge field is coupled with a spinless matter field in one spatial dimension. These models have the twofold advantage of a careful control over finite-$n$ effects, and of providing a {\color{red} controlled} approximation of lattice U$(1)$ quantum electrodynamics in the large-$n$ limit. 
We will focus on the ground state properties for different values of $n$ and other parameters of the model. In particular, we will find that, with no background field (corresponding, in the $n\to\infty$ limit, to the $CP$-invariant point of the Schwinger model), the system undergoes a quantum phase transition towards regions of the parameter space in which the parity and charge conjugation symmetries are spontaneously broken. The transition will be characterized in terms of finite-size scaling and critical exponents, which will fall into the Ising universality class. Such quantum phase transition does not survive the introduction of a background field, since in this case the excitations become always massive. We will numerically study the $\mathbb{Z}_n$-models for $n=2$ to $n=8$ and extrapolate the results to the large $n$-limit, showing that the corresponding known phase transition of the U$(1)$-model is recovered. The features of this class of transitions and its asymptotic properties will be carefully scrutinized.

The article is organized as follows. In Section~\ref{sec:discrete}, we introduce the massive Schwinger model in $1+1$ dimensions and discuss the paths to discretization of space and gauge degrees of freedom. Section~\ref{sec:zn} includes the definition of $\mathbb{Z}_n$ gauge models and the presentation of their general features and scaling properties. In Section~\ref{sec:z3} we study in detail the case $n=3$, characterizing its ground state properties and the quantum phase transition at a negative critical mass, in absence of background field. Section~\ref{sec:zlarger3} is devoted to a presentation of the results obtained in all cases $n=2 \div 8$, $n\neq3$, focusing on the different phenomenology of the even and odd cases, while the  details of the numerical results of all these cases are given in the Appendix. In Section~\ref{sec:compara} we summarize our results and recover the U$(1)$ model in the  limit of large $n$. In section~\ref{sec:coldatom}, we comment on a possible implementation of the proposed class of models in a cold atomic platform. We finally draw our conclusions in Section~\ref{sec:concl}.

\section{Discretization of one-dimensional QED}
\label{sec:discrete}

Quantum Electrodynamics in one spatial dimension is a U$(1)$ gauge theory, describing the interaction of a charged particle (``electron''), represented by a spinor field $\psi(t,x)$, and the electromagnetic field $F_{\mu\nu}=\partial_{\mu}A_{\nu}-\partial_{\nu}A_{\mu}$, associated to the potential $A_{\mu}$, with $\mu,\nu=0,1$. The classical Lagrangian density is determined by the minimal coupling prescription:
\begin{equation}
\mathcal{L} = \psi^{\dagger}\gamma^0 \left[\gamma^{\mu}(\ii\partial_{\mu} + g A_{\mu} ) - m \right] \psi - \frac{1}{4}F_{\mu\nu}F^{\mu\nu} ,
\end{equation}
where $m$ and $g$ are the electron mass and charge, respectively, and $\{\gamma^{\mu},\gamma^{\nu}\}=2\eta^{\mu\nu}$ with $\eta=\mathrm{diag}(1,-1)$. The properties of the theory are strongly characterized by the absence of transverse degrees of freedom: the electron, described by a two-component spinor, is spinless, and the only independent component of the electromagnetic tensor is the electric field $E=F_{01}$. While quantization of the spinor field is determined by the canonical equal-time anticommutators
\begin{align}
\{\psi(t,x),\psi(t,x')\} & =0, \\ \{\psi(t,x),\psi^{\dagger}(t,x')\} & =\delta(x-x'),
\end{align}
a gauge choice is necessary to quantize the electromagnetic potential. In the canonical gauge, the temporal component $A_0$ is set to zero, while the spatial component $A:=A_1$ is taken as the conjugate variable to $E$:
\begin{equation}
\label{EAcont}
[E(t,x),A(t,x')]=\ii\delta(x-x').
\end{equation}
This choice, leading to the Hamiltonian
\begin{equation}
H = \int \d x \, \left\{ \psi^{\dagger}\gamma^0 \left[ -\gamma^1 (\ii\partial_1 + g A) + m \right] \psi + \frac{E^2}{2} \right\} ,
\label{eq:Schwinger}
\end{equation}
does not allow one to enforce  Gauss' law $G(x) = 0$, with
\begin{equation}
G(x) = \partial_1 E(x) - g \psi^{\dagger}(x) \psi(x),
\end{equation}
as an operator constraint. However, since $[G(x),G(x')]=0$ and $[G(x),H]=0$ due to~\eqref{EAcont}, it is possible to select the physical subspace of states $\ket{\psi}$ for which $G(x)\ket{\psi}=0$, which will be denoted by
\begin{equation}
G(x) \approx 0,
\end{equation}
at all space points. 

\begin{figure}
\centering
\includegraphics[width=0.4\textwidth]{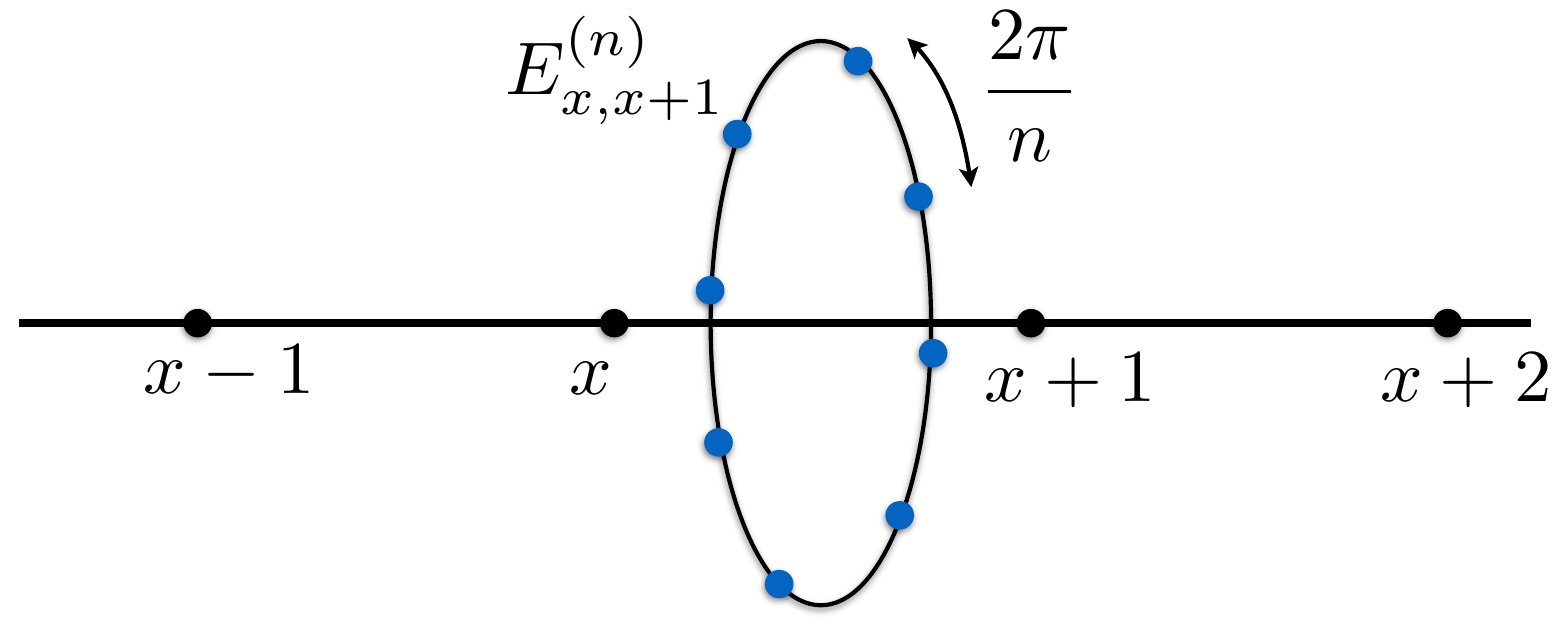}
\caption{Discretization of the Schwinger model. Fermionic matter lives on sites $x \in \mathbb{Z}$. Electric field $E^{(n)}_{x,x+1}$ lives on links between adjacent sites and takes discrete values,  
$e_{k}=\sqrt{\frac{2\pi}{n}} (k -\frac{n -1}{2})$, with $k\in \mathbb{Z}_n=\{0,1,\cdots,n-1\}$. }
\label{fig:zn} 
\end{figure}

In the following, we will consider two kinds of discretization, towards classical and quantum simulations of the model. The first one is spatial discretization: the continuum model will be replaced by an approximation on a linear lattice of points with spacing $a$, making the continuous space variable $x \in \mathbb{R}$ discrete: $x \in \mathbb{Z}$. The second one is the approximation of the  gauge group U$(1)$ with a finite group, which is essential if one wants to work with a finite number of local degrees of freedom in the gauge variables. 
This can be done essentially in two ways, based on the generalization of the commutation relation $[E,U]=\eta U$, \textcolor{red}{where $E$ and $A$ are two conjugated operators ($[E,A]=\ii$),} $U=\e^{-\ii\eta A}$  \textcolor{red}{is} the gauge comparator \textcolor{red}{and $\eta\in\mathbb{R}$ is a constant with the same dimensions as $E$}. One option is to focus on the preservation of the above commutator. This is the approach taken, for example, in quantum link models~\cite{qlm1,qlm2,qlm3,qlm4wiese}, in which the operators $E$ and $U$ are replaced with spin variables. Another option is to require that the \emph{group} commutator $\e^{\ii\xi E} \e^{-\ii\eta A} = \e^{\ii\eta \xi} \e^{-\ii\eta A} \e^{\ii\xi E}$, which is equivalent to the previous one in the U$(1)$ case, be satisfied by unitary operators for discrete values of $\eta$ and $\xi$~\cite{NEFMPP}. We will follow the latter strategy, that entails the reduction of gauge invariance to a finite group $\mathbb{Z}_n$. A pictorial representation of the gauge degrees of freedom is shown in Fig.~\ref{fig:zn}.

As emphasized in the Introduction, we thus obtain an \emph{exact} finite implementation of the gauge group commutator at the Hamiltonian level. What is yet to be understood is the proper scaling of the dynamical variables in order that the $\mathbb{Z}_n$ model correctly reproduce a lattice U$(1)$ model, with a continuous electric field. Clearly, $n$ must go to $+\infty$. However, proper scaling will be necessary for $E=E^{(n)}$. As we shall see, this scaling is dictated by Schwinger himself~\cite{schwinger2001quantum}.

\section{Lattice $\mathbb{Z}_n$-QED model}
\label{sec:zn}

\textcolor{red}{The Schwinger model~\eqref{eq:Schwinger} described in the previous section can be discretized on a one-dimensional lattice. For convenience, we shall first redefine, the vector potential $A\to A/g$ and the electric field $E\to g E$. With this transformation, that leaves the commutation relation~\eqref{EAcont} unchanged, the charge is absorbed in the minimal coupling, to reappear in the energy density of the free electric field. Correspondingly, we redefine the parameters in the one parameters groups $U(\eta),V(\xi)$ by $\eta\rightarrow \eta/g, \xi\rightarrow g\xi$.}

\textcolor{red}{The lattice Hamiltonian reads}~\cite{montvay1997quantum,qlm4wiese,melnikov}
\begin{align}
\label{HamQED}
H = & - \frac{1}{2a} \sum_x \!\left(\psi^{\dagger}_x
U_{x,x+1}\psi_{x+1}+\mathrm{H.c.} \right)\! \nonumber \\
& + m \sum_x(-1)^x\psi_x^{\dagger}\psi_x+ \frac{g^2a}{2}\sum_x
E_{x,x+1}^2,
\end{align}
with $x$ labelling the sites of a one-dimensional lattice of spacing $a$. Here:\\
i) fermionic matter is represented by one-component spinor {\color{red}{creation/annihilation operators $\psi_{x}^\dagger,\psi_{x}$, defined on each site $x$, so that $\sum_x \psi_{x}^\dagger\psi_{x}$ represents the total number of fermions in the system}}; \\
ii) a staggered mass $(-1)^x m$ is introduced, so that the positive- and negative-mass components of the Dirac spinor live respectively at even and odd lattice sites, avoiding in this way the fermion-doubling problem~\cite{kogut1975hamiltonian,susskind1977lattice,rothe1992lattice,montvay1997quantum}; \\
iii) gauge fields are defined on the links $(x,x+1)$ of the lattice through the pair of variables $E_{x,x+1}$ (electric field) and $A_{x,x+1}$ (vector potential) or equivalently the comparators $U_{x,x+1}(\eta)= \e^{-\ii \eta A_{x,x+1} }$ and $V_{x,x+1}(\xi)= \e^{+\ii \xi E_{x,x+1} }$, commuting at different sites, and satisfying the group canonical commutation relations: 
\begin{equation}
\label{cr}
V_{x,x+1}(\xi)\, U_{x,x+1}(\eta) = \e^{\ii\eta
\xi} U_{x,x+1}(\eta)\  V_{x,x+1}(\xi), 
\end{equation}
for $\xi,\eta\in\mathbb{R}$, which are  equivalent exponentiated versions of the (algebra) canonical commutation relations $[E_{x,x+1},A_{x,x+1} ] = \ii $.\\
\textcolor{red}{Notice that in~(\ref{HamQED}), the coupling constants $m$ and $g$ are the same as in the continuum}\textcolor{red}{, both with the dimensions of energy, while all the field operators are dimensionless.}

\textcolor{red}{In analogy to what is done in fermionic lattice models that are usually used to describe condensed matter systems and for reasons that will become clear in the next section, in the following we will actually use a slightly modified Hamiltonian by introducing an additional dimensionless parameters $t$ in front of the first, kinetic, term so to write the dimensionless Hamiltonian
\begin{align}
\label{HamQED2}
h_t & = \frac{2}{g^2 a}H_t =  - \frac{t}{g^2a^2} \sum_x \!\left(\psi^{\dagger}_x
U_{x,x+1}\psi_{x+1}+\mathrm{H.c.} \right) \nonumber \\
& + \frac{2m}{g^2 a} \sum_x(-1)^x\psi_x^{\dagger}\psi_x+ \sum_x
E_{x,x+1}^2 \, ,
\end{align}
proportional to the Hamiltonian density.  
We can also see from this expression that the coefficient $g^2a/2$ fixes the scale of the mass, while the coefficient $g^2a^2$ the one of the parameter $t$. Thus, the numerical simulations described in Sections~\ref{sec:z3} and~\ref{sec:zlarger3} will be performed by setting $g^2a/2=1$ and $g^2a^2=1$. The standard lattice Schwinger model is then recovered for $t=1$.}

In addition, the theory must respect  Gauss' law, encoding the U$(1)$  gauge symmetry of the model, that reads
\begin{equation}
\label{Gx}
 G_x \equiv \psi_x^{\dagger}\psi_x + \frac{1}{2}[(-1)^x-1] -(E_{x,x+1}-E_{x-1,x}) \approx 0. 
\end{equation}
The Hamiltonian~(\ref{HamQED}) is invariant under $C$ and $P$ symmetries that for staggered fermions read (assuming an infinite lattice or setting $-x\equiv 2L-x$ for a finite lattice with an even number $N=2L$ of sites, labeled from $0$ to $2L-1$):
\begin{equation}
P\, : \, \left\{ \begin{array}{ll}
\psi_x \rightarrow \psi _{-x}, & \psi_x^\dagger \rightarrow \psi _{-x}^\dagger, \\
E_{x,x+1} \rightarrow E_{-(x+1),-x}, & U_{x,x+1} \rightarrow U_{-(x+1),-x} ^\dagger,
\end{array}  \right. 
\label{parity} 
\end{equation}
\begin{equation}
C\, : \, \left\{ \begin{array}{ll}
\psi_x \rightarrow (-1)^{x+1}  \psi _{x+1}^\dagger, & \psi_{x}^\dagger \rightarrow (-1)^{x+1} \psi _{x+1}, \\
E_{x,x+1} \rightarrow - E_{x+1,x+2}, & U_{x,x+1} \rightarrow U_{x+1,x+2} ^\dagger.
\end{array}  \right.  \label{charge} 
\end{equation}

The Hamiltonian~(\ref{HamQED}) can be approximated via a discrete Abelian $\mathbb{Z}_n$-gauge model~\cite{NEFMPP}, that can be obtained from a 
{\it finite dimensional} representation of the two-parameter projective unitary Weyl~\cite{weyl1950theory} group $\left\{ \e^{\ii(\xi E_{x,x+1}-\eta A_{x,x+1})}\right\}_{\xi,\eta \in \mathbb{R}}$. \\
For the two particular cases $(\xi,\eta)=(0,\sqrt{2\pi/n})$  and $(\xi,\eta)=(\sqrt{2\pi/n},0)$ one gets the two operators $U_{x,x+1} =\e^{-\ii \sqrt{\frac{2\pi}{n}}A_{x,x+1} }$ and $V_{x,x+1}= \e^{\ii \sqrt{\frac{2\pi}{n}} E_{x,x+1} }$, that satisfy the commutation relations
\begin{equation}\label{commSW}
U_{x,x+1}^{\ell} V_{x,x+1}^k = \e^{\ii\frac{2\pi}{n}k\ell}\, V_{x,x+1}^{k} U_{x,x+1}^{\ell} \quad
\mbox{with } k,\ell\in \mathbb{Z}_n.
\end{equation}
which is a discrete $\mathbb{Z}_n$ version of~(\ref{cr}).
This representation can be implemented by considering an $n$-dimensional Hilbert space ${\cal H}_n$ defined on each link, and choosing an orthonormal basis $\{ |v_k \rangle \}_{0\leq k\leq n-1}$.   
Dropping the link index, we consider the diagonal operator $V$ acting as
\begin{equation}
V\ket{v_k} = \e^{-\ii 2 \pi k/n} \ket{v_k}.
\end{equation}
The operator $U$ is instead defined as that operator that performs a cyclic permutation of the basis states:
\begin{equation}
\label{cyclic}
U |v_k\rangle = |v_{k+1}\rangle \quad \mbox{for } k<n-1, \quad
U|v_{n-1}\rangle = |v_{0}\rangle .
\end{equation}
Some simple algebra shows that these operators do indeed satisfy the Schwinger-Weyl commutation relations~\eqref{commSW}.
Let us remark that this representation exactly implements the unitarity of both operators.

Thus the dynamics of the $\mathbb{Z}_n$-model is determined by the Hamiltonian~(\ref{HamQED}), 
where the discrete version of the electric field $E_{x,x+1}$ is given by the Hermitian operator that is diagonal in the $\{ |v_k \rangle \}$ basis, with eigenvalues  
\begin{equation}
\label{eigenvn}
e_k = \sqrt{\frac{2\pi}{n}}\left(k-\frac{n-1}{2}+\phi\right). 
\end{equation}
{\textcolor{red}{In all cases, the eigenvalues of the electric field are symmetric around zero, with a maximum value $E_{\mathrm{max}}=\sqrt{2\pi/n}(n-1+\phi)/2$. We notice that, for $\phi=0$, it is possible to have zero electric field only if $n$ is odd.} }A value $\phi\neq 0$ corresponds to adding a background field that can be obtained by placing charges at the boundaries of the chain, thus yielding different charge sectors, that are known to be super-selected. {\textcolor{red}{It is indeed known that this model displays $\theta$-vacua~\cite{coleman2} , which can be related to the axial anomaly~\cite{manton} via the spectral flow of the Hamiltonian operator (when imposing periodic boundary conditions) or to unusual twisted boundary conditions for the fermionic field~\cite{wipf}.  In this case the $P$ and $C$ symmetries are explicitly broken. In the case of even $n$, the minimum eigenvalues~\eqref{eigenvn} are doubly degenerate for $\phi=0$. As a consequence, in the strong-coupling limit, in which the $U$-dependent terms in~\eqref{HamQED} are neglected, the energetic cost of creating a fermion-antifermion pair from the vacuum vanishes: this feature is typical of theories with $\theta=\pi$~\cite{coleman2}.}}

\section{Scaling properties of the Hamiltonian}
\label{sec:scaling}

{\textcolor{red}{Before starting to numerically investigate the Hamiltonian, some comments are in order to establish the correctness of Eq.~(\ref{HamQED2}) to suitability represent a quantum simulator for 1-dimensional QED. 
Being in particular interested in its critical properties, our analysis needs to contain a careful check of the scaling properties of the discretized Hamiltonian as we change the different parameters that appear in it, including the $a\rightarrow 0$ limit (continuum limit), the $N\rightarrow \infty$ limit (infinite volume limit), the $n\rightarrow \infty$ limit ($U(1)$-limit). Close to a critical point, at which physical constants and observables are functionally related by universal laws, it is very hard to control these different cases independently, both from an analytical and a numerical point of view. However, we can resort to well-known techniques based on a finite-size scaling analysis guided by universal scaling properties. We have chosen to perform this study in two steps: first, we consider a particular $\mathbb{Z}_n$-model, by keeping $n$ fixed, and perform a finite-size scaling in the dimension of the spatial lattice; second, we let $n$ increase and analyze the large-$n$ limit. }}

{\textcolor{red}{To this end, we first notice that the parameter $t$ can be used to understand critical properties of the model. Let us suppose that the system undergoes a phase transition for a critical value of the mass, $m_c(t)$, which may depend on $t$. At this particular point the dimensonless Hamiltonian~(\ref{HamQED2})  should be scale invariant. The coefficient $t$ can be absorbed in a re-scaling of the lattice spacing, $a\rightarrow \tilde{a} = a/\sqrt{t}$, and:
\begin{align}\label{HamQED3}
h_c = & 
- \frac{1}{g^2\tilde{a}^2} \sum_x \!\left(\psi^{\dagger}_x
U_{x,x+1}\psi_{x+1}+\mathrm{H.c.} \right) \nonumber \\
& + \frac{2m_c(t)}{\sqrt{t}\, g^2\tilde{a}} \sum_x(-1)^x\psi_x^{\dagger}\psi_x+ \sum_x
E_{x,x+1}^2 \, ,
\end{align}
In a mean-field approach, in which possible anomalous dimensions of the field are neglected, the coefficient in front of the second addend must be independent of $t$. In other words, the critical value of the mass scales like
\begin{equation}\label{mt0}
m_c(t) = \alpha \sqrt{t} ,
\end{equation}
where $\alpha\equiv m_c(t=1)$. 
We will examine accurately how the critical value of the mass depends on $t$ in the numerical simulations of the next sections, where we will  see that its behaviour does not significantly deviates  from the one predicted here. Therefore we will obtain the continuum limit critical mass $m_c$ by setting: $m_c=m_c(t=1)=\alpha$.}}

{\textcolor{red}{Incidentally, let us remark that the limit $t\rightarrow \infty$ is not equivalent to the limit $a\rightarrow 0$, since the coefficient $t$ weights the kinetic term differently with respect to the mass and the electric energy terms; in particular, in the $t=0$ case we recover a classical limit which can be exactly solved, while in the large $t\rightarrow \infty$ limit we deal with a pure kinetic Hamiltonian which cannot display any phase transition. }}

\textcolor{red}{Second, we want to study the large-$n$ limit. It is important to notice that, as shown in~\cite{NEFMPP}, the scaling of the eigenvalues of the electric field with $n$ as given in Eq.~(\ref{eigenvn}) is fixed by requiring that the U$(1)$-limit is recovered when $n\rightarrow+\infty$. Also, recalling that two consecutive values of the electric field differ by $\sqrt{2\pi/n}$, it is convenient to collect such a factor and work with the dimensionless Hamiltonian
\begin{align}\label{HamQED4}
h^{(n)}_t = & \frac{2}{g_n^2 a}H^{(n)}_t =  
-  \frac{t}{g_n^2a^2} \sum_x \!\left(\psi^{\dagger}_x
U_{x,x+1}\psi_{x+1}+\mathrm{H.c.} \right) \nonumber \\
&  +\frac{2m}{g_n^2a} \sum_x(-1)^x\psi_x^{\dagger}\psi_x+ \sum_x
\tilde{E}_{x,x+1}^2 \, ,
\end{align}
where now $\tilde{E}_{x,x+1}$ has eigenvalues $(k-(n-1)/2+\phi)$ with unit spacing (hence, independent of $n$), and 
\begin{equation}
\label{eq:gndef}
g_n=g \sqrt{2\pi/n}. 
\end{equation}
In the same spirit as before, we can conclude that now (with a slight abuse of notation)}
\begin{equation}\label{mtn}
 m_c(t) = \alpha_n \sqrt{t} .
\end{equation}
Comparing the Hamiltonian density~(\ref{HamQED4}) to~(\ref{HamQED3}) in the limit $n\to\infty$, we can conclude that the asymptotic value of $\alpha_n g/g_n$ must approach the coefficient $\alpha$ appearing in Eq.~(\ref{mt0}), namely
\begin{equation}\label{mtn}
\alpha= \lim_{n\rightarrow \infty} \alpha_n \sqrt{n/2\pi} .
\end{equation}
{\textcolor{red}{In the following, to perform numerical calculations, we will consider the Hamiltonian~(\ref{HamQED4}) defined on a lattice of size $N=2L$ with  open boundary conditions. We will work in the sector with one fermion for each ``physical site'', i.e. with $N_{\textrm{part}}=N/2=L$ particles. Also, as explained above, we will set $g^2 a/2=1$ (which sets the units of energy) and $g^2 a^2=1$ (which sets the units of $t$). }}

{\textcolor{red}{We will first present the $\mathbb{Z}_3$-model, in order to illustrate all the details of our treatment. We will then discuss the general $\mathbb{Z}_n$-model, for both odd and even $n$. As we will see, these two cases need to be considered separately.}}

\section{Lattice $\mathbb{Z}_3$-QED model}
\label{sec:z3}

\subsection{Hilbert space and gauge-invariant subspace}
\label{sec:zaa}

\begin{figure}
\centering
\subfigure[\label{fig:fermions}]{\includegraphics[width=0.45\textwidth]{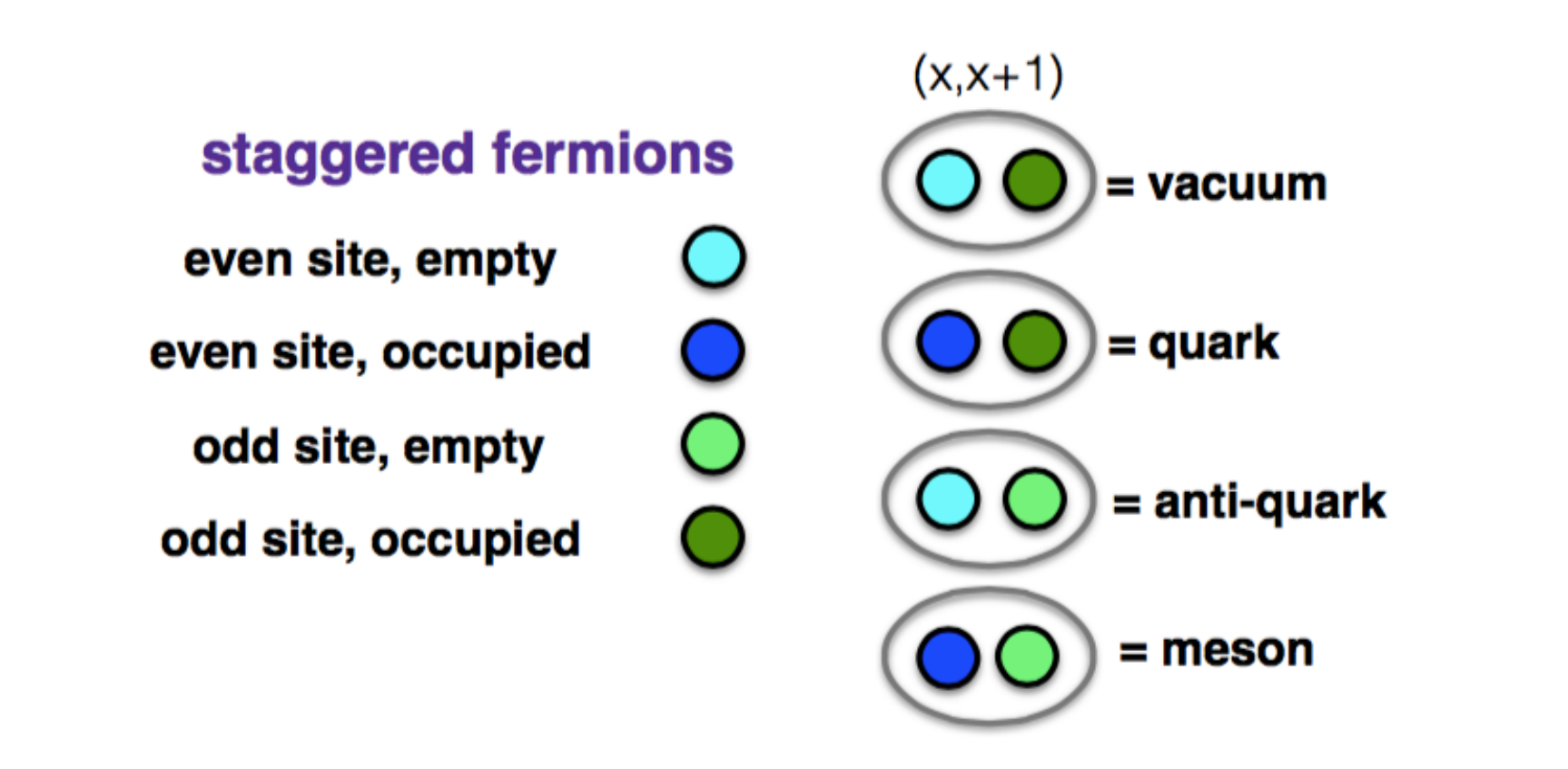}}
\subfigure[\label{fig:fields}]{\includegraphics[width=0.25\textwidth]{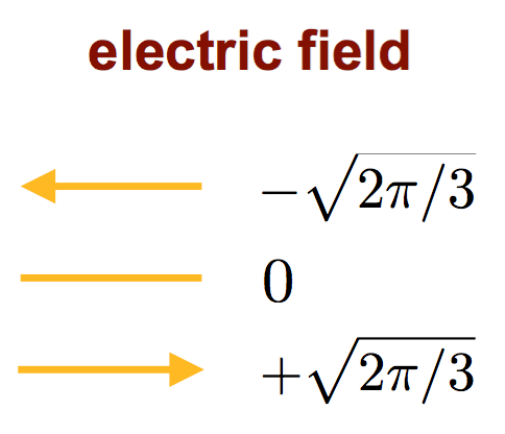}}
\caption{$\mathbb{Z}_3$-model. Local Hilbert space for (a) staggered fermions; (b) electric field.}
\end{figure}

As mentioned in Sec.~\ref{sec:zn}, in the Schwinger model each ``physical fermion'' is represented by a pair of staggered fermions sitting in nearby sites, with even/odd sites occupied by positive/negative mass particles. Thus the vacuum state (Dirac sea) is obtained by leaving the even sites empty and occupying the odd ones. The presence/absence of a fermion in an even/odd site is interpreted as the presence of a quark/anti-quark, while a meson is a configuration made up of a  quark and an anti-quark. This is shown in Fig.~\ref{fig:fermions}. On each link $(x,x+1)$, the electric field can only assume one of  the three values $E= \sqrt{2\pi/3} (k-1)$, with $k \in \mathbb{Z}_3=\{0,1, 2\}$, which will be represented as an arrow pointing left, an un-oriented segment, and an arrow pointing left, respectively, as shown in Fig.~\ref{fig:fields}. 

\begin{figure}
\centering
\subfigure[\label{fig:gauge}]{\includegraphics[scale=0.25]{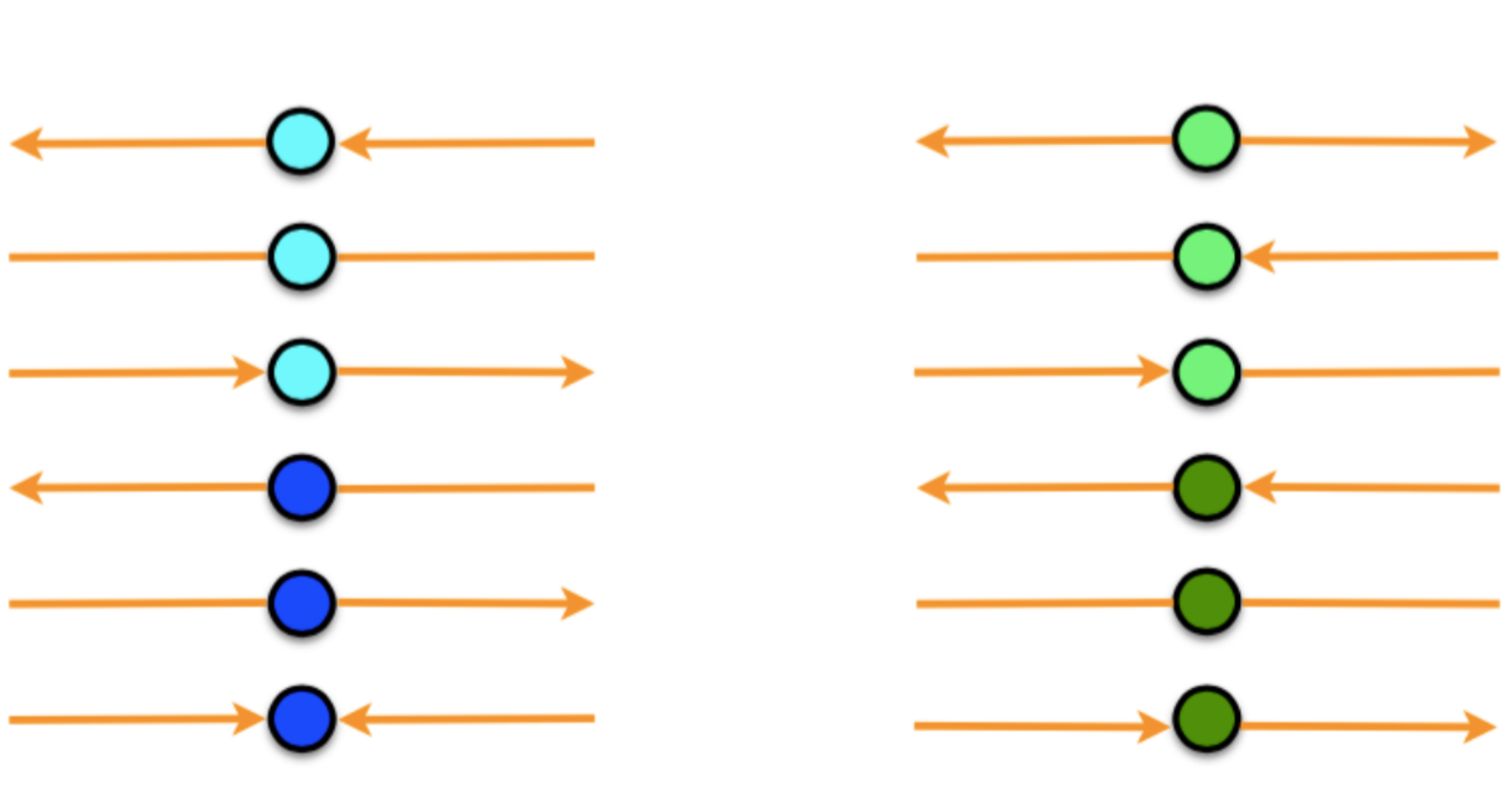}}
\subfigure[\label{fig:hilbert}]{\includegraphics[scale=0.45]{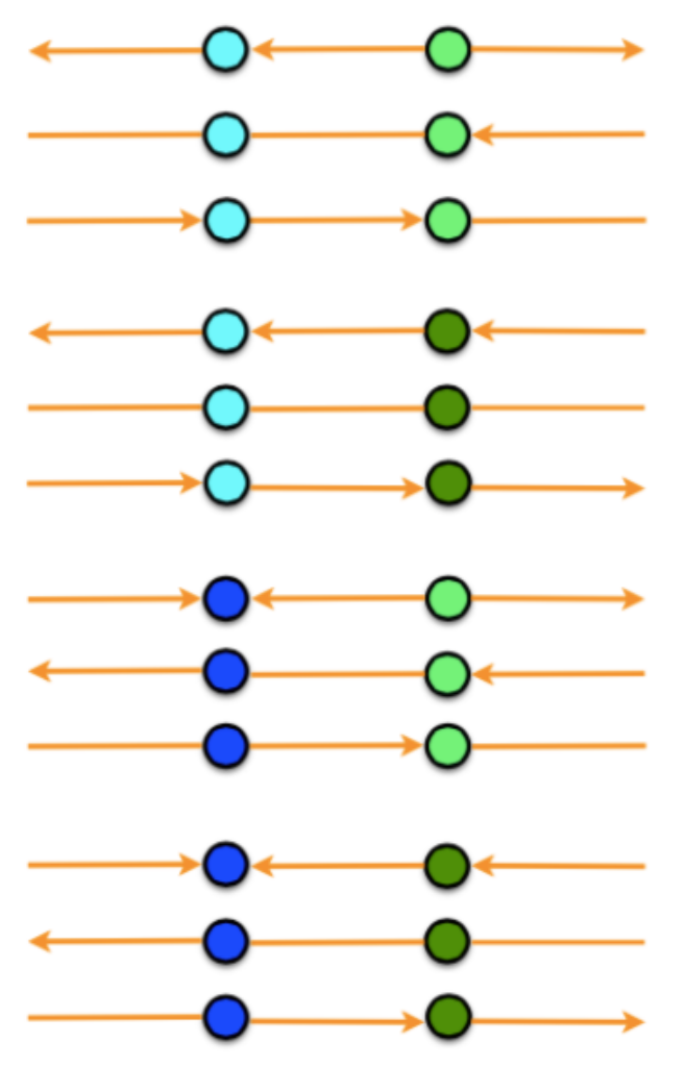}}
\caption{$\mathbb{Z}_3$-model. (a) Gauge-invariant Hilbert space associated with even/odd sites; (b) Gauge-invariant Hilbert space associated with a pair of even/odd sites, i.e.\ a ``physical site''.}
\label{fig:hspace}
\end{figure}

Thus the total Hilbert space associated with an even site, together with its two adjacent links, is $2\times3\times3=18$ dimensional. But  Gauss' law forces the physical states to belong to an invariant subspace which is constructed out of those states for which the value of the electric field $k_r$ at the right link is either $k_r=k_l$, if the site is empty, or $k_r=k_l+1(\mathrm{mod} \,3)$  if the site between links is occupied by a (positive mass) fermion. The situation is similar for odd sites, for which  Gauss' law constrains physical states to have either $k_r=k_l-1(\mathrm{mod}\, 3)$, if the site between links is empty, or $k_r=k_l$ if the site is occupied by a (negative mass) fermion. This is displayed in Fig.~\ref{fig:gauge}. Notice that we have $2\times3=6$ independent configurations for each site. The gauge invariant states of a ``physical site'' are obtained by gluing together an even and an odd site that share a common value for the electric field in the link between them, obtaining $2\times 6=12$ possible configurations, as shown in Fig.~\ref{fig:hilbert}. 

\begin{figure}
\centering
\includegraphics[scale=0.3]{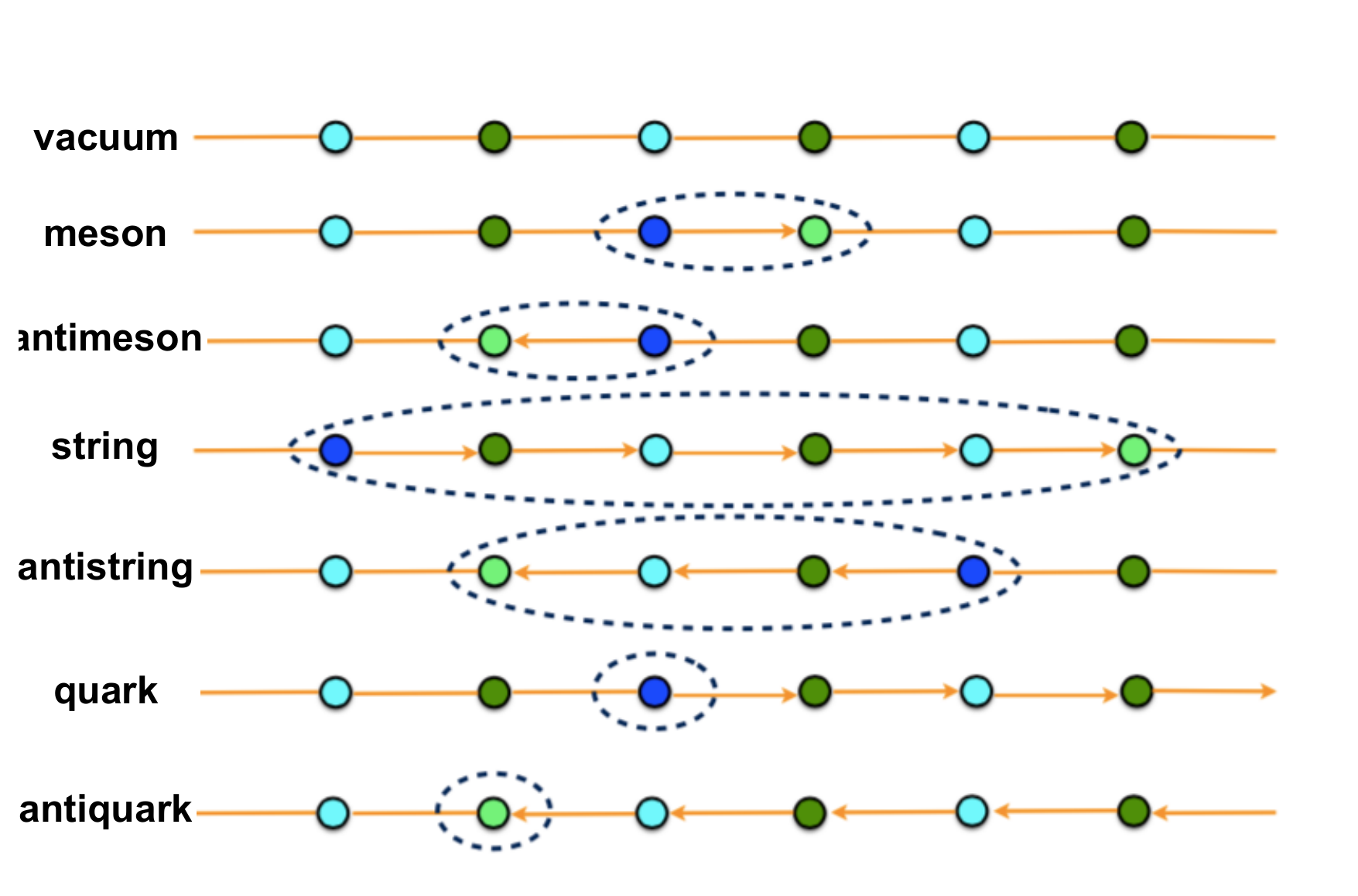}
\caption{$\mathbb{Z}_3$-model. Some notable gauge invariant configurations.}\label{fig:4} 
\end{figure}

It is easy to see that, for a chain with $N$ sites (with open boundary conditions), the dimension of the gauge-invariant subspace is $2^{N}\times 3$. Some notable examples of gauge-invariant states in a chain are shown in Fig.~\ref{fig:4}. 

\begin{figure}
\centering
\subfigure[\label{fig:vacuum}]{\includegraphics[width=0.45\textwidth]{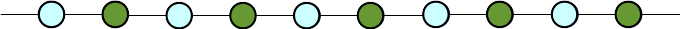}}
\subfigure[\label{fig:mesons}]{\includegraphics[width=0.45\textwidth]{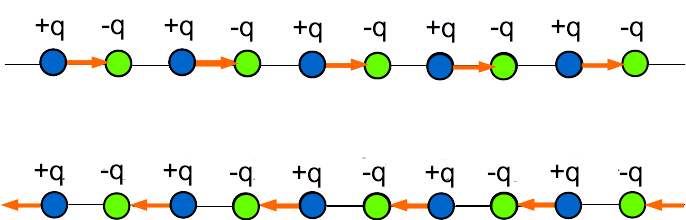}}
\caption{$\mathbb{Z}_3$-model. (a) Dirac sea; (b) Mesonic (top) and antimesonic (bottom) states.}
\end{figure}

The ground state of the Hamiltonian~(\ref{HamQED}) will be given by the completely filled Dirac sea (see Fig.~\ref{fig:vacuum}) for large positive $m$, while for large negative $m$ the system will tend to choose between the two states shown in Fig.~\ref{fig:mesons}, where mesons/antimesons have formed. Notice that the Dirac sea is invariant under both parity and charge conjugation, while $P$ and $C$ map the mesonic and antimesonic states into each other. 
These two cases are clearly distinguished by the mean value of the electric field operator 
\begin{equation}
\varSigma=\frac{1}{N}\sum_{x}\left\langle E_{x,x+1}\right\rangle 
\end{equation}
that we will use as a kind of order parameter, since it vanishes for the Dirac sea and takes the values $\pm \pi/3$ for the mesonic/antimesonic states. An abrupt change of this quantity signals the existence of a possible phase transition, whose existence has to be confirmed by looking at the appropriate scaling of observables and thermodynamical quantities. 

\subsection{Numerical investigation of the critical point with no background field}
\label{sec:zbb}

We perform numerical calculations by means of a finite-size DMRG code~\cite{DMRG2}  in which gauge invariance is exactly enforced. This is obtained by using a unit cell given by a pair of nearby (even and odd) sites, whose local Hilbert space is the span of the 12 gauge-invariant states described in Fig.~\ref{fig:hspace}. This is implemented at each step of the algorithm, with a twofold advantage: decreasing the computing time by working in a restricted space, and avoiding transitions out of the gauge-invariant subspace.
We work with up  to $N=80$ sites ($L=40$ pairs), while keeping $1000$ DMRG states at most.  These values are large enough to ensure stability of our findings and small errors. 

\begin{figure}
\centering
\subfigure[\label{fig:sigma3_a}]{\includegraphics[width=0.5\textwidth]{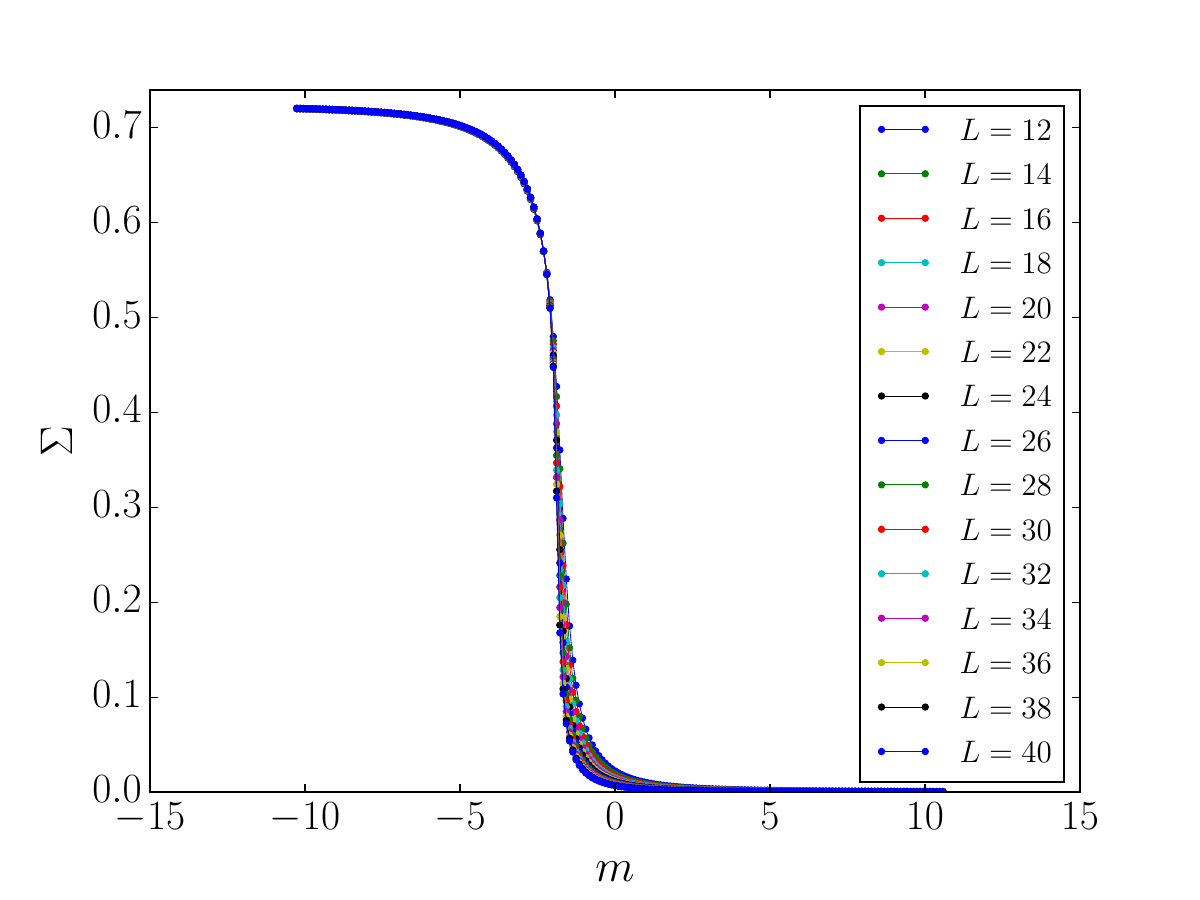}}
\subfigure[\label{fig:sigma3_b}]{\includegraphics[width=0.5\textwidth]{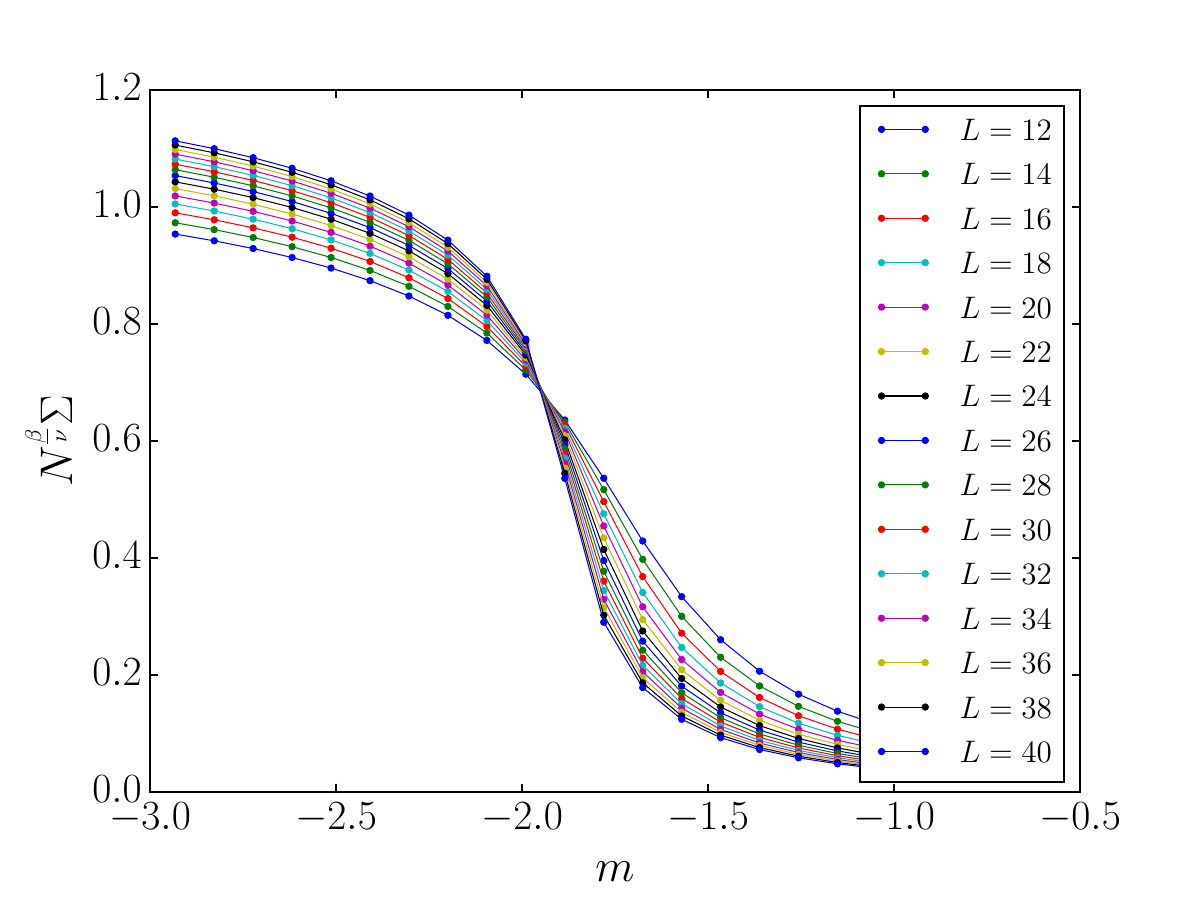}}
\caption{$\mathbb{Z}_3$-model. (a) Order parameter $\varSigma$ as a function of $m$, for different system size $L$; (b) Same plot as in (a), in the vicinity of the phase transition.}
\end{figure}

We first numerically study the Hamiltonian~(\ref{HamQED4}) at the $CP$-invariant point, i.e.\ in absence of a background field: $\tilde{E}_{x,x+1}\in\{-1,0,+1\}$. We start by choosing $t=2\pi/3$ so as to work with the operator
\begin{align} \label{h3}
{\color{red} h^{(3)}_{t=\frac{2\pi}{3}} } = & - \sum_x \!\left(\psi^{\dagger}_x U_{x,x+1}\psi_{x+1}+\mathrm{H.c.} \right)\! \nonumber \\ & 
+ \frac{3}{2\pi} m \sum_x(-1)^x\psi_x^{\dagger}\psi_x+  \sum_x \tilde{E}_{x,x+1}^2.
\end{align}
\textcolor{red}{Notice that, here and in the following sections, mass is expressed in units of $g^2 a/2$.} The behavior of the observable $\varSigma$ as a function of $m$ is displayed in Fig.~\ref{fig:sigma3_a} for different system sizes, ranging from $L=12$ to $L=40$. We see that, as expected, $\Sigma$ essentially vanishes at large positive $m$ and tends to the value $\sqrt{2\pi/3}/2\simeq0.724$ for large negative $m$. 

\begin{figure}
\centering
\includegraphics[width=0.5\textwidth]{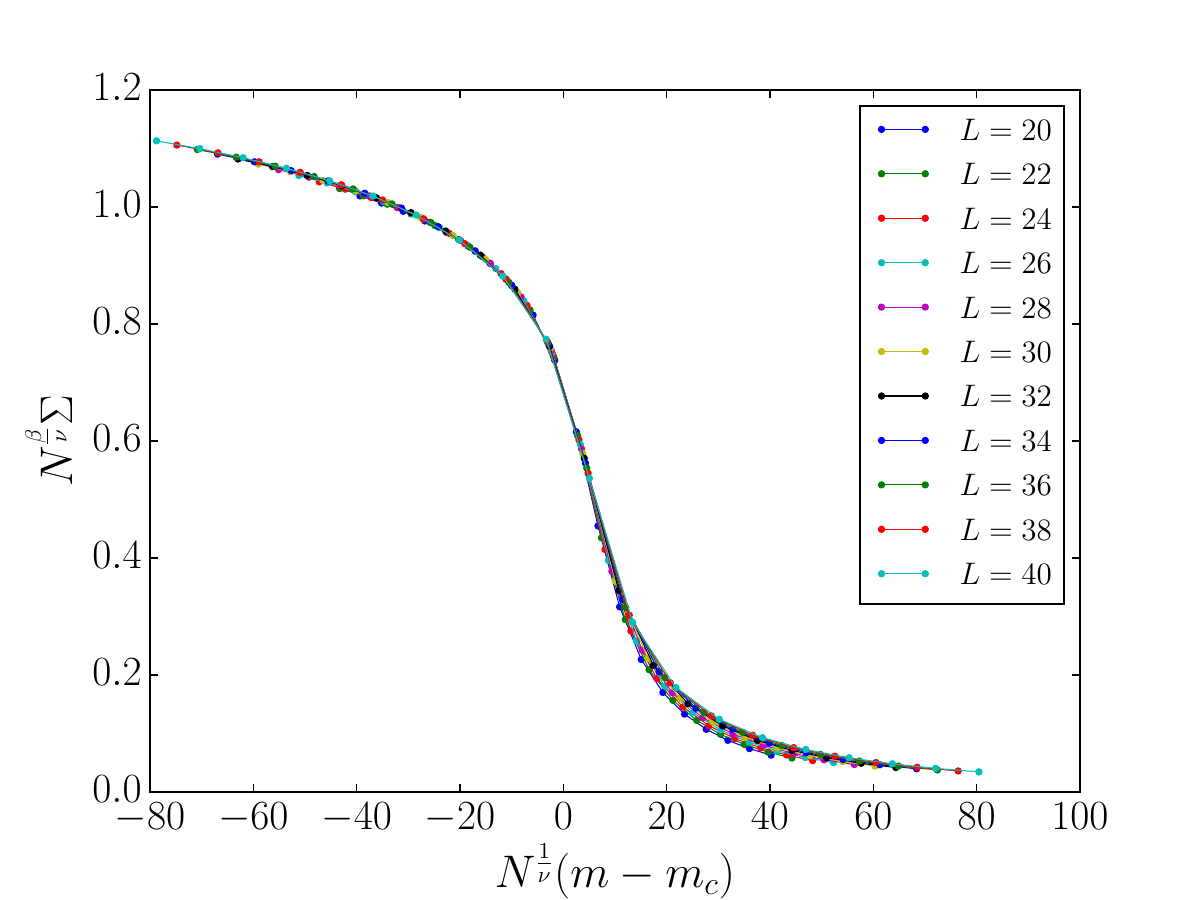}
\caption{$\mathbb{Z}_3$-model.  Scaling of $\lambda(x)$ for different system size $L$.}\label{fig:collapse} 
\end{figure}

In Fig.~\ref{fig:sigma3_b} we zoom on the central region, showing a steeper transition as the system size increases. This strongly suggests that we are in presence of a phase transition, at a critical value of the mass which corresponds to the point where all curves intersect. We can estimate this value if we make a hypothesis about the nature of the phase transition: indeed, if we know the critical exponents, we can calculate $m_c$ by using the fact that $\varSigma$ should scale with the system size $N$ according to the finite-size scaling formula~\cite{Cardy}
\begin{equation}
\varSigma=N^{-\frac{\beta}{\nu}}\lambda \big( N^{\frac{1}{\nu}}(m-m_{c})\big)\label{eq:scaling_parametro_ordine}
\end{equation}
where $\lambda$ is a universal function.
By taking into account suggestions from the continuum limit~\cite{coleman2} and the symmetries of the model, we anticipate that the phase transition is of the Ising-type, so that $\beta=1/8$ and $\nu=1$. A fit of the data yields then $m_c = -1.948 \pm 0.025$, where the error has been estimated as the semi-interval between the numerical points. 
We now have to look at the numerical curves given by $N^{\frac{\beta}{\nu}}\varSigma$ versus $N^{\frac{1}{\nu}}(m-m_{c})$, for different $N$, which should all collapse onto the same universal curve $\lambda(x)$. This behaviour is clearly seen in Fig.~\ref{fig:collapse}.

\begin{figure}
\centering
\includegraphics[width=0.5\textwidth]{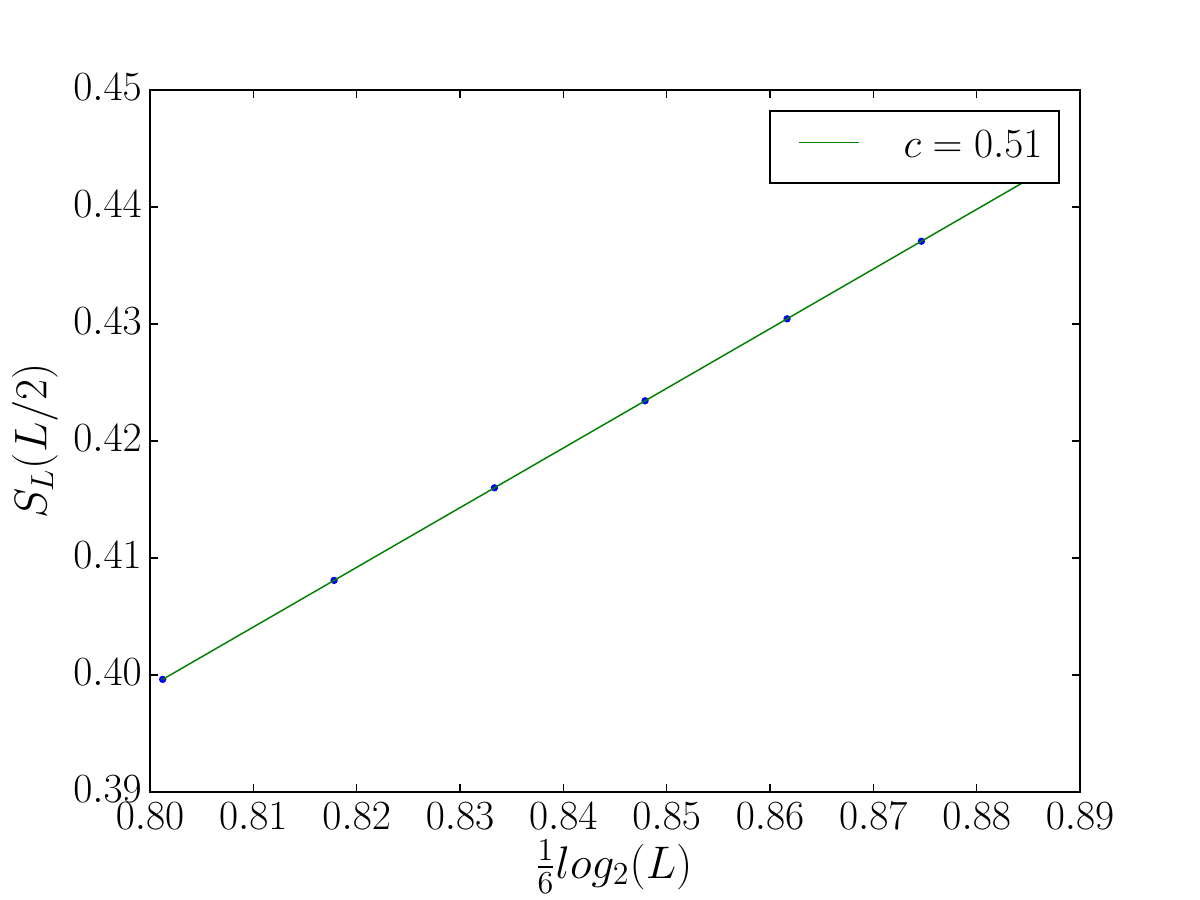}
\caption{$\mathbb{Z}_3$-model. Entanglement entropy $S_{L}(L/2)$ versus system size $L$.}\label{fig:entropy} 
\end{figure}

Once we have an estimate for the critical mass, we can confirm the validity of our hypothesis by calculating other quantities.
Figure \ref{fig:entropy} displays the entanglement entropy of a subsystem of size $L/2$ at the critical point, which---according to conformal field theory~\cite{cardy_calabrese}---should scale logarithmically with the system size according to the law
\begin{equation}
S_{L}\left(\frac{L}{2}\right)=\frac{c}{6} \log_{2}(L)+s_0,
\end{equation} 
where $s_0$ is a constant (which can depend on the boundary conditions and other details of the model) while $c$ is the central charge.
The fit yields $c =  0.51 \pm 0.01$, in perfect agreement with the central charge of the Ising model, $c=1/2$. 

\begin{figure}
\subfigure[\label{fig:delta}]{\includegraphics[width=0.5\textwidth]{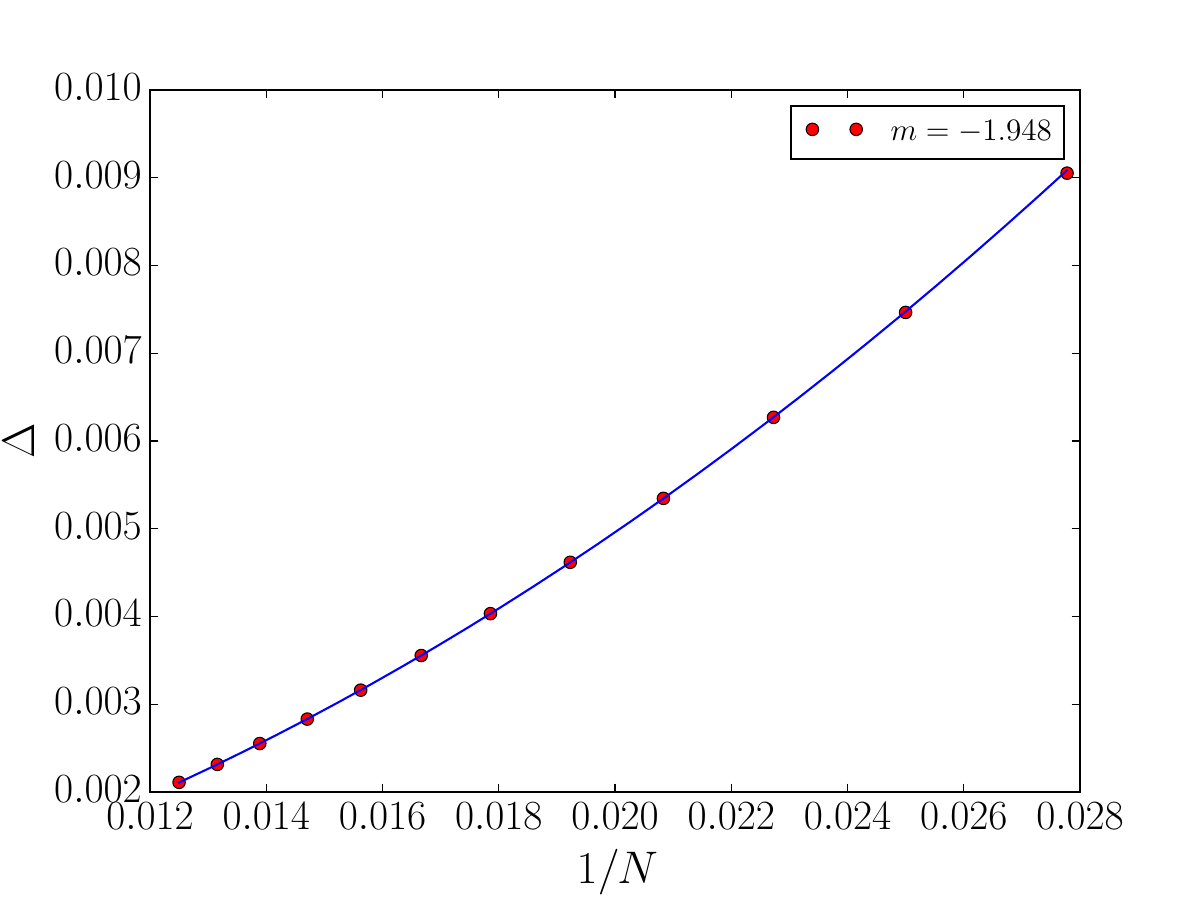}}
\subfigure[\label{fig:gamma}]{\includegraphics[width=0.5\textwidth]{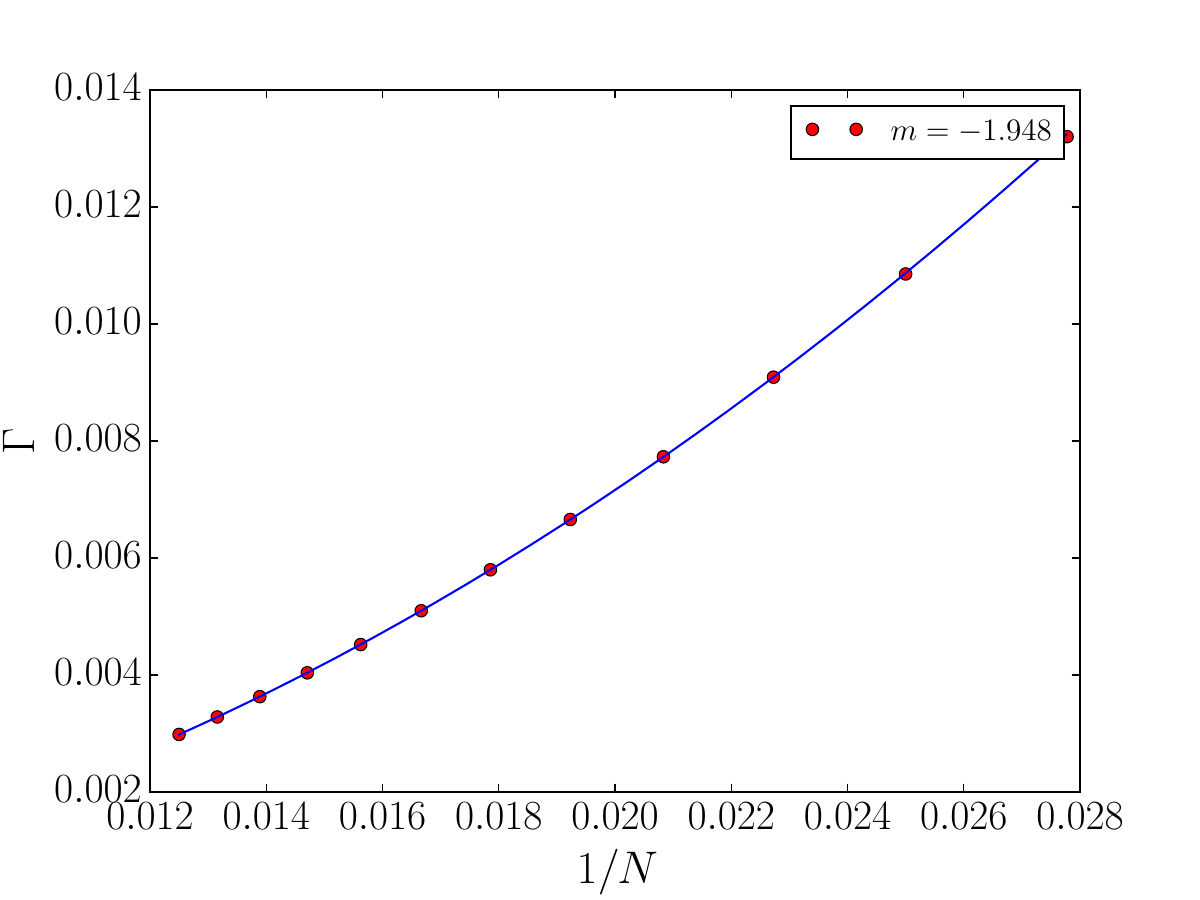}}
\caption{$\mathbb{Z}_3$-model. (a) Gap $\Delta$ of the first excited state versus system size $N=2L$; fit according to Eq.~(\ref{eq:delta}).
(b) Gap $\Gamma$ of the second excited state versus system size $N=2L$; fit according to Eq.~(\ref{eq:gamma}).}
\end{figure}

Additional information can be obtained by looking at the scaling of excited states with the size of the system: Figures \ref{fig:delta} and~\ref{fig:gamma} display the behaviour of the gaps $\Delta$ and $\Gamma$ of the first two excited states, which---again according to conformal field theory~\cite{henkel}---should obey
\begin{eqnarray}
&\Delta=\varepsilon_{1}(N)-\varepsilon_{0}=\frac{\pi v_{s}x_{s}}{N^{2}}, \label{eq:delta}\\
&\Gamma=\varepsilon_{2}(N)-\varepsilon_{0}=\frac{\pi v_{s}(x_{s}+1)}{N^{2}}, \label{eq:gamma}
\end{eqnarray}
$\varepsilon_0$ being the ground state energy density, $v_s$ a speed, and $x_{s}$ the surface critical exponent, which is equal to 2 for the Ising model with open boundary conditions. We numerically find
\begin{equation}
\frac{\Delta}{\Gamma}=\frac{x_{s}}{x_{s}+1}=0.6671\pm0.0008 \Rightarrow x_{s}=2.004\pm0.007.
\end{equation}
Plugging this result back into Eq.~(\ref{eq:delta}) we can  also estimate the speed $v_s$, obtaining 
\begin{equation}
v_s = 1.56\pm0.08
\end{equation} 
(a number  very close to $\pi/2$).
We remark that surface exponents are found for states that can be obtained from the ground state by changing from periodic to anti-periodic boundary conditions. The system here shows spontaneous symmetry breaking to one of the two degenerate polarized states and thus the first excited state can be represented as a kink-like solution which interpolates between these two degenerate polarized minima, in agreement with what is found in the continuum~\cite{coleman2}. These results are also fully compatible with recent~\cite{buyens, shimizu_kuramashi} and less recent~\cite{byrnes2002} numerical results. Interestingly, similar conclusions hold also at finite temperature~\cite{sachs_wipf}.

For $t=0$, the system undergoes a (first order) phase transition between the Dirac sea, depicted in Fig.\ \ref{fig:vacuum}, with an energy per pair $E_{\textrm{Dirac}}/L = -m$,
and the mesonic state in Fig.\ \ref{fig:mesons}, with energy per pair $E_{\textrm{meson}}/L = m+ 2\pi/3$.
The critical value of the mass $m_0^{(3)}=-\pi/3 \simeq -1.047$ is obtained at $E_{\mathrm{meson}}=E_{\mathrm{Dirac}}$. 

In order to test the validity of Eq.~(\ref{mtn}), we have repeated this procedure for several values of $t$, checking that the Ising transition is always present and calculating numerically $m_c(t)$. 
Our numerical findings for $m_c(t)$ as a function of $t$, as well as other useful information, are given in Appendix \ref{addinfo}. 

A numerical fit of the form
\begin{equation}
m_c(t) = m_0^{(n)} + \alpha_n \sqrt{t} + \beta_n t
\label{formfit}
\end{equation} 
yields the values
\begin{align}
&m_{0}^{(3)}=-1.0472\pm0.0001, \\
 &\alpha_3=-0.603\pm0.001,\\
&\beta_3=-0.02\pm0.01 .
\end{align} 
Let us notice that, as expected from the predicted behaviour~(\ref{mtn}), the coefficient of the linear term is much smaller than the one of the square-root term, thus yielding a negligible correction, at least for not too large values of $t$. This will also be apparent in Fig.\  \ref{fig:comparazione_diagrammi} (green points and green fitting curve).

\subsection{Numerical investigation of the critical point in presence of a background field}

It is known~\cite{coleman2} that the Schwinger model should exhibit a phase transition only at the $CP$-invariant point. In order to check if this is the case also in our model, we have scrutinized the effects of a constant background field. We present here just one representative example, by considering an electric field 
\begin{equation}
\tilde{E}_{x,x+1} = k+1/3, \qquad k\in\{-1,0,+1\}
\end{equation}  
in the Hamiltonian~(\ref{h3}).

\begin{figure}
\subfigure[\label{fig:sigma3backa}]{\includegraphics[width=0.5\textwidth]{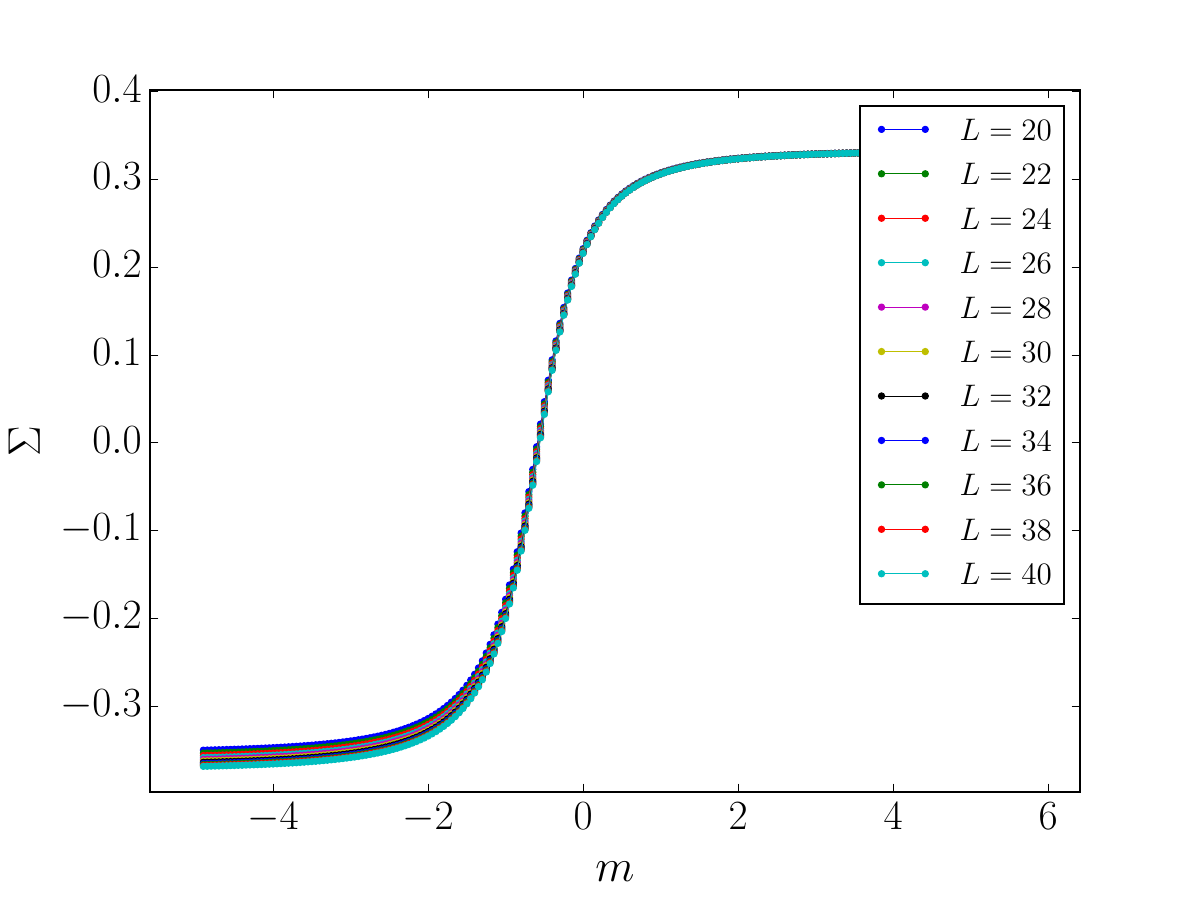}}
\subfigure[\label{fig:sigma3backb}]{\includegraphics[width=0.5\textwidth]{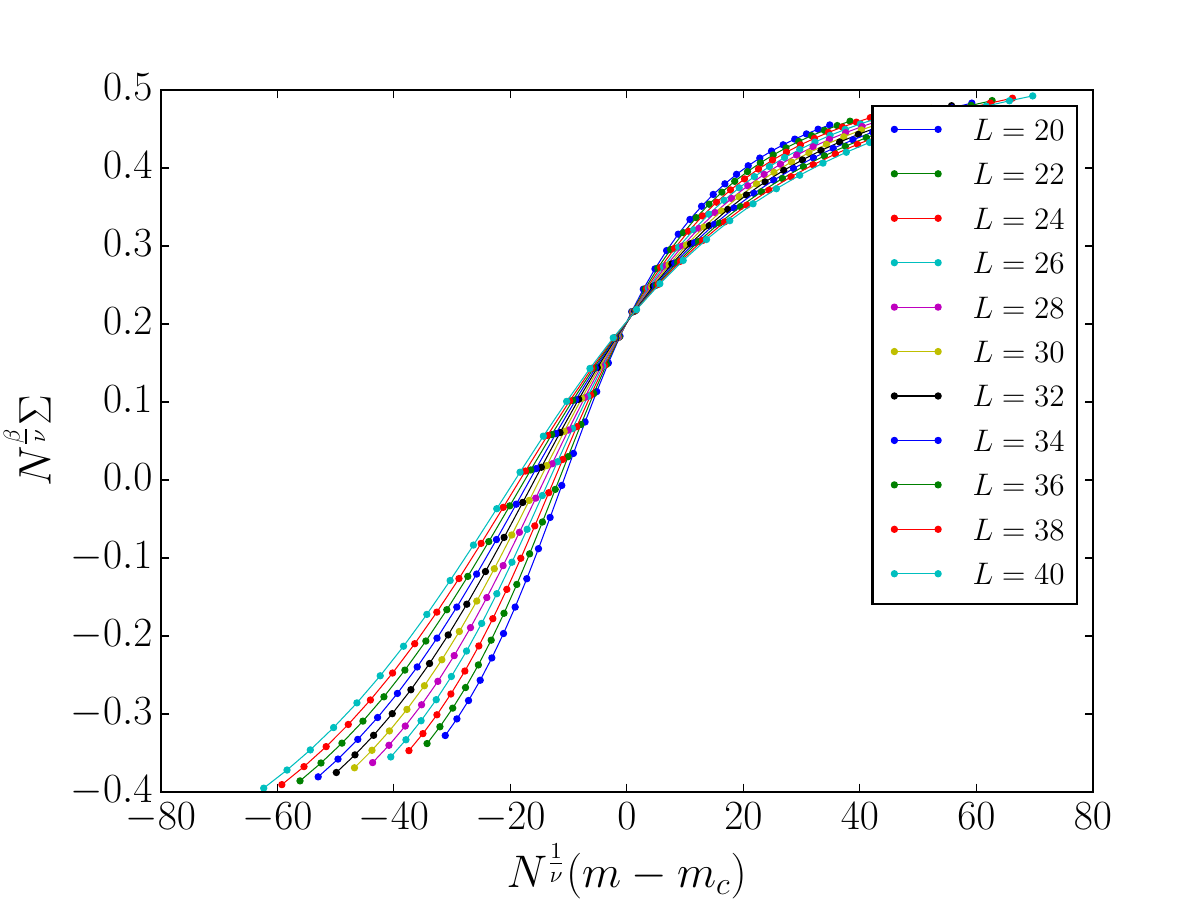}}
\caption{$\mathbb{Z}_3$-model. (a) Observable $\Sigma$ as a function of $m$ in presence of a background electric field; (b) Non-universal scaling of the function $\lambda(x)$.}
\end{figure}

\begin{figure}
\subfigure[\label{fig:entropy3backa}]{\includegraphics[width=0.5\textwidth]{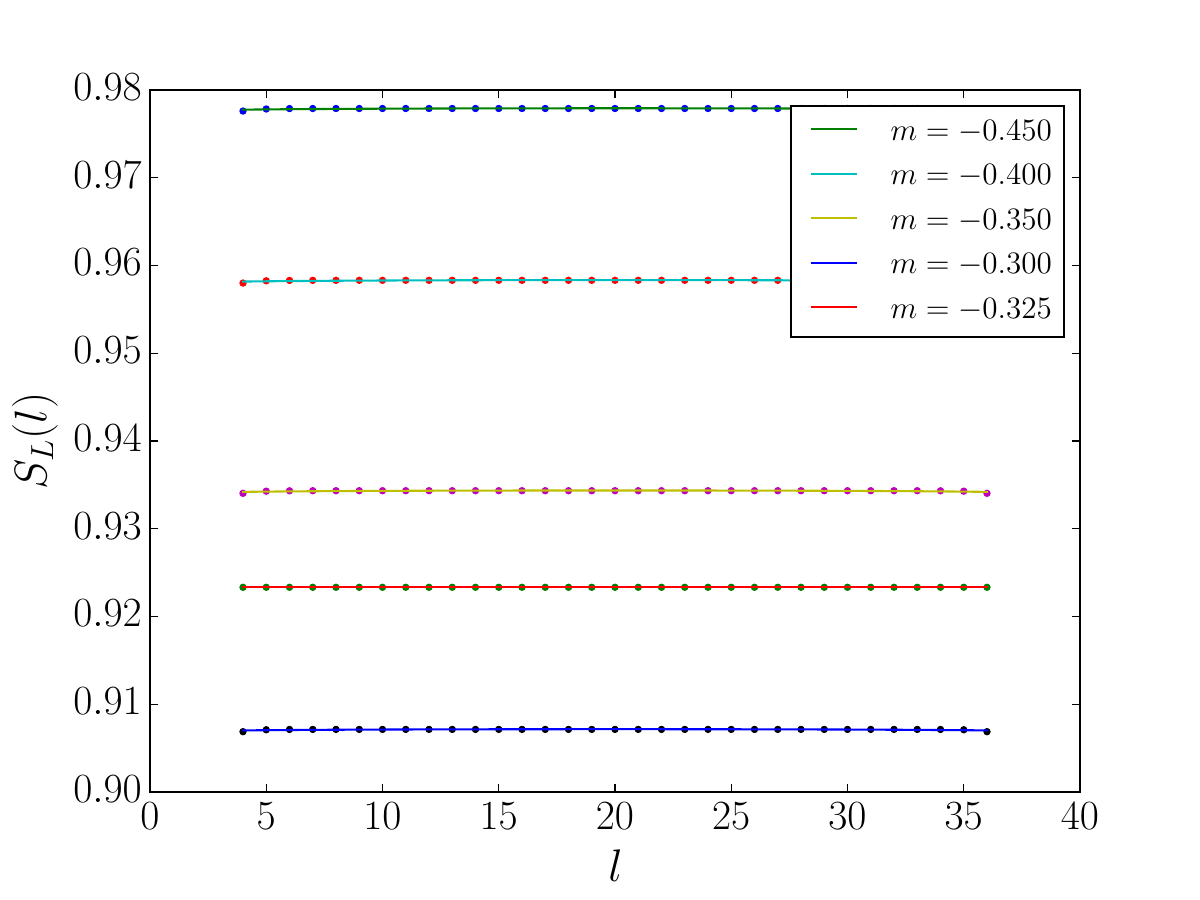}}
\subfigure[\label{fig:delta3backb}]{\includegraphics[width=0.5\textwidth]{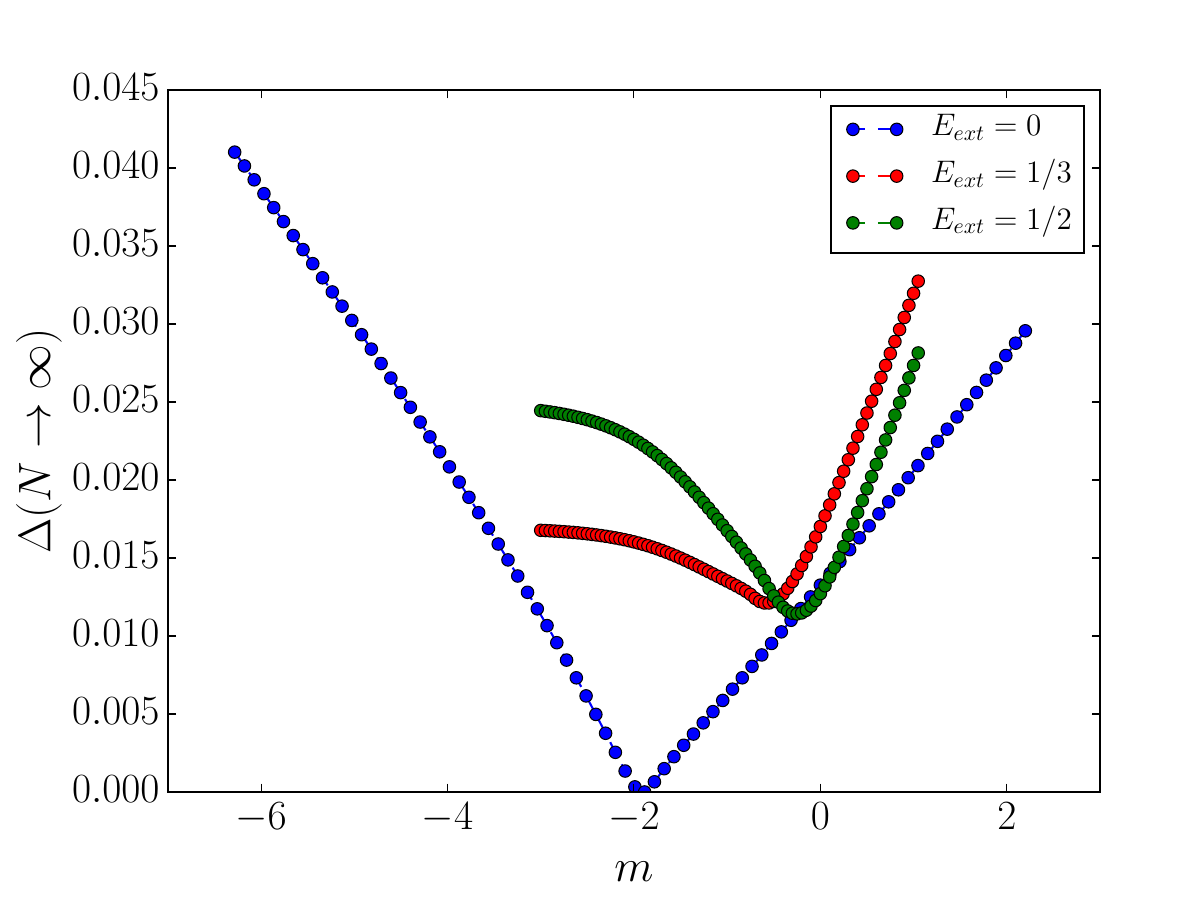}}
\caption{$\mathbb{Z}_3$-model. (a) Entanglement entropy $S_L(l)$ of an interval of size $l$, in presence of a background electric field; (b) Gap $\Delta$ with (red and green dots) and without background field (blue dots). }
\end{figure}

As shown in Fig.~\ref{fig:sigma3backa}, the observable $\varSigma$ still shows a very sharp transition between a negative and a positive value.  But we are now in presence of a cross-over, rather than a phase transition, as it can be inferred by performing a scaling analysis. Indeed, the function $\lambda$ in Eq.~(\ref{eq:scaling_parametro_ordine}) changes for different system size $N$ and does not have a universal character, as one can infer from Fig.~\ref{fig:sigma3backb}. Also, the entanglement entropy $S_L(l)$ does not scale with the size $l$ of the interval, as predicted by conformal field theory~\cite{cardy_calabrese}, but is rather constant, except for some small edge effects, also at the crossing point $m^*=-0.325$ (see Fig.~\ref{fig:entropy3backa}).  These results suggest that, in presence of a background field, the gap never closes, as the numerics confirms (see Fig.~\ref{fig:delta3backb}, red dots). 

Finally, we also checked the case of a background electric field which is halfway between two integer values: 
\begin{equation}
\tilde{E}_{x,x+1} = k+1/2, \qquad k\in\{-1,0,+1\}.
\end{equation} 
Also in this case the model is gapped for any value of the mass, as shown in Fig.~\ref{fig:delta3backb} (green dots). 
At a first sight, this situations looks very similar to the case of even $n$ with no background field, when the possible spectrum of the electric field does not include zero, being still invariant under a sign change. However, the two cases are very different, since, as we will discuss in the next section, the $\mathbb{Z}_n$-model with even $n$  and no background field, which is $CP$-invariant, still exhibits a phase transition.


\section{Lattice $\mathbb{Z}_n$-QED model for other values of $n$}
\label{sec:zlarger3}

The analysis performed in the previous section for the $\mathbb{Z}_3$-model can be repeated for all values of $n$. In the following we will consider only the case with no background field.
The dimension of the gauge-invariant Hilbert subspace for a chain with $N$ sites is $2^N \times n$, increasing only linearly with $n$, since the electric field can now take the $n$ values $- \sqrt{\frac{2\pi}{n}} \frac{n-1}{2},\cdots,  +\sqrt{\frac{2\pi}{n}}\frac{n-1}{2}$. Similarly to what was done for the $\mathbb{Z}_3$ case, we rescale the electric field as: $E_{x,x+1} = \sqrt{2\pi/n} \tilde{E}_{x,x+1} $ and study the Hamiltonian~(\ref{HamQED4}), with $g^2a^2=g^2a/2=1$. 
One must consider odd and even $n$ separately.

\subsection{Odd $n$}

As for the case $n=3$ presented in the previous section, if $n$ is odd we anticipate a phase transition from a phase where the ground state is the Dirac sea for large positive $m$, to a phase in which the ground state is a meson/antimeson state for large negative $m$. At $t=0$, there is a first-order phase transition between these two states, at a critical mass $m_0^{(n)}$ that can be easily found by comparing the energy of these two states, given respectivley by $E_{\mathrm{Dirac}}/L = -m$ and $E_{\mathrm{meson}}/L = m+ 2\pi/n$, thus yielding the critical value 
\begin{equation}
m_0^{(n)}=-\pi/n. 
\end{equation}

\begin{figure}
\centering
\includegraphics[width=0.5\textwidth]{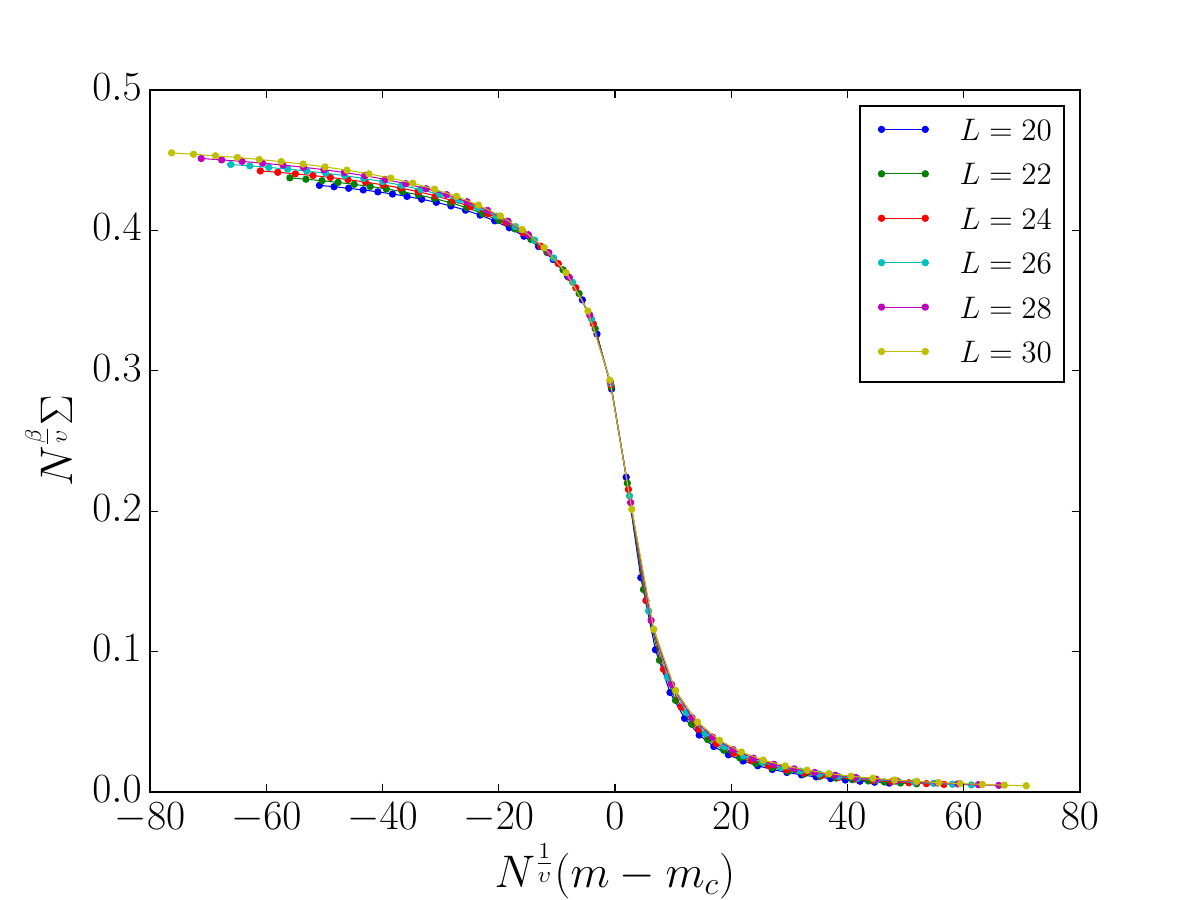}
\caption{$\mathbb{Z}_5$-model. Universal scaling function $\lambda(x)$ close to the transition point
($t=2 \pi/5$). 
}
\label{fig:order5} 
\end{figure}

Fot $t\neq0$ we must resort to our DMRG code and perform an analysis identical to the one presented for the $\mathbb{Z}_3$-model, assuming again that the phase transition falls in the Ising universality class.
As an example, in Fig.~\ref{fig:order5} we show the behaviour of the function $\lambda$ of Eq.~(\ref{eq:scaling_parametro_ordine}) for  different system size in the $\mathbb{Z}_5$-model and for $t=2\pi/5$, proving its universality in this case as well.
We have performed an exhaustive analysis of the $\mathbb{Z}_5$- and $\mathbb{Z}_7$-models, obtaining the value of $m_c(t)$ as function of $t$ in both cases, as summarized in the Appendix and in Fig.\ \ref{fig:comparazione_diagrammi}.

We can now fit these data with the formula~(\ref{formfit}) to get an estimate of the coefficients $\alpha_n$, $\beta_n$ with $n=5$ and $7$. The numerical results are summarized in Table~\ref{table:msum} and show an excellent agreement with the theoretically predicted value $m_0^{(n)}= -\pi/n$. Also, as for the $n=3$ case,  the coefficient of the linear term is much smaller than the one of the square-root term.

\subsection{Even $n$}

Even-$n$ models are different from odd-$n$ ones since the electric field cannot take the value zero, still being $CP$-invariant. This means that we are working in a different super-selection sector corresponding to a different total charge at the boundary.

\begin{figure}
\centering
\subfigure[\label{fig:base_z2}]{\includegraphics[width=0.5\textwidth]{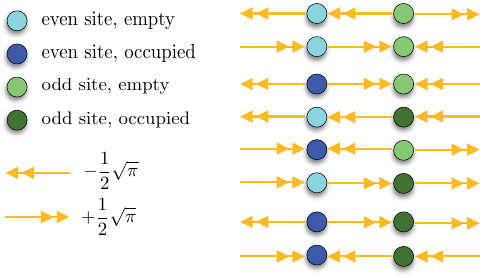}}
\subfigure[\label{fig:ground_states_z2}]{\includegraphics[width=0.5\textwidth]{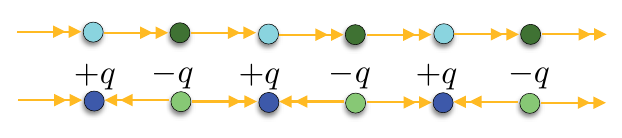}}
\caption{$\mathbb{Z}_{2}$-model. (a) Basis of the gauge-invariant Hilbert subspace; (b) ground states for large positive/negative $m$.}
\label{fig:z2_complessivo}
\end{figure}

Let us first consider the case $n=2$. This is a very small (in fact, the smallest non-trivial) value, and one may expect some peculiarities, due to the fact that the scaling arguments in Eqs.~(\ref{HamQED3})-(\ref{mt0}) do not apply.
\textcolor{red}{Actually, since the electric field Hamiltonian becomes trivial for $\phi=0$, the presence of a phase transition only depends on the ratio of the coefficients of the hopping term and the mass term. Hence, the critical mass would be linear in $t$.} 
The gauge-invariant Hilbert subspace for a pair of sites is $8$-dimensional and a basis is shown in Fig.~\ref{fig:base_z2}. The electric field can assume the two values $-\sqrt{\pi}/2,+\sqrt{\pi}/2$. The transition is from a phase where the ground state is the uniformly polarized vacuum, for large positive $m$, to a ground state in which the electric field has alternating signs on links, for large negative $m$. These states are shown in Fig.~\ref{fig:ground_states_z2}.

\begin{figure}
\centering
\subfigure[\label{fig:scaling_par_ordine1_z2}]{\includegraphics[width=0.5\textwidth]{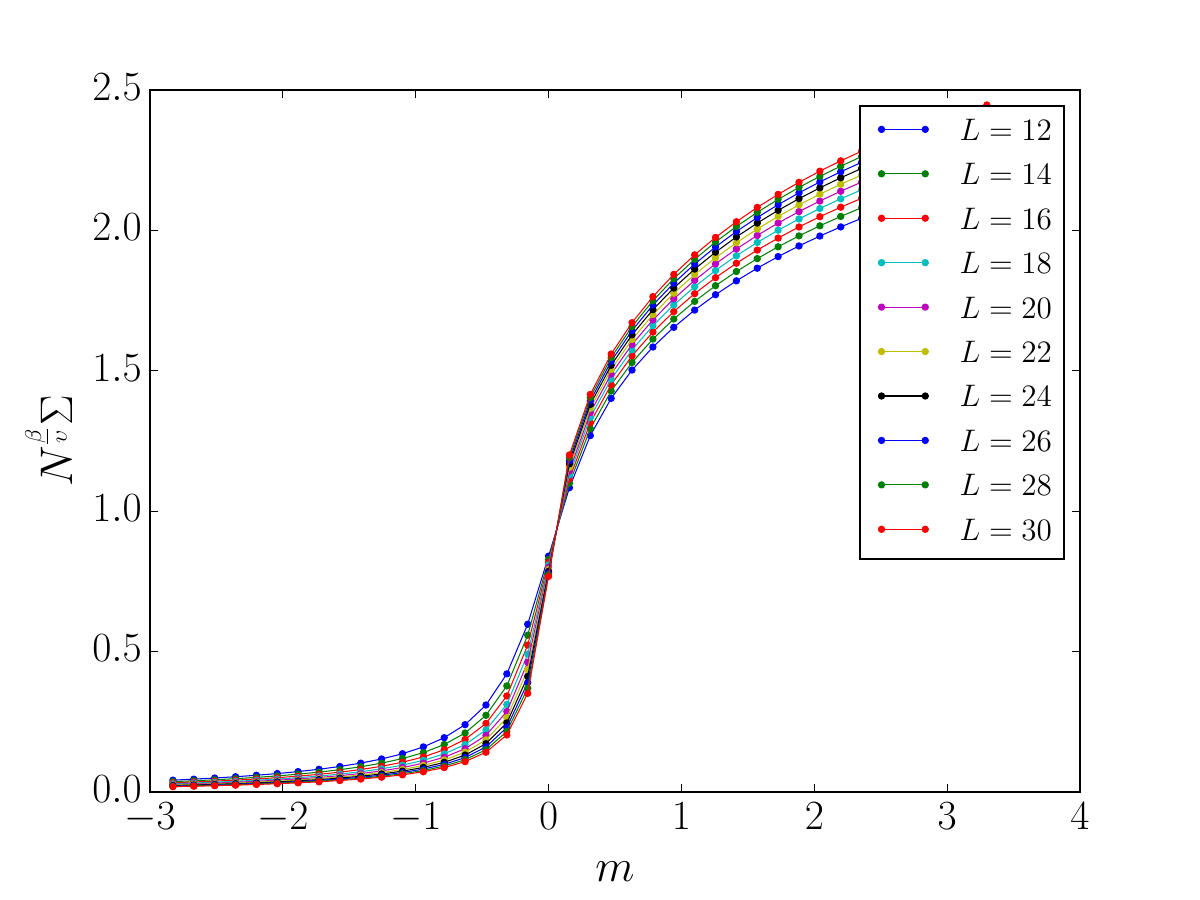}}
\subfigure[\label{fig:scaling_par_ordine2_z2}]{\includegraphics[width=0.5\textwidth]{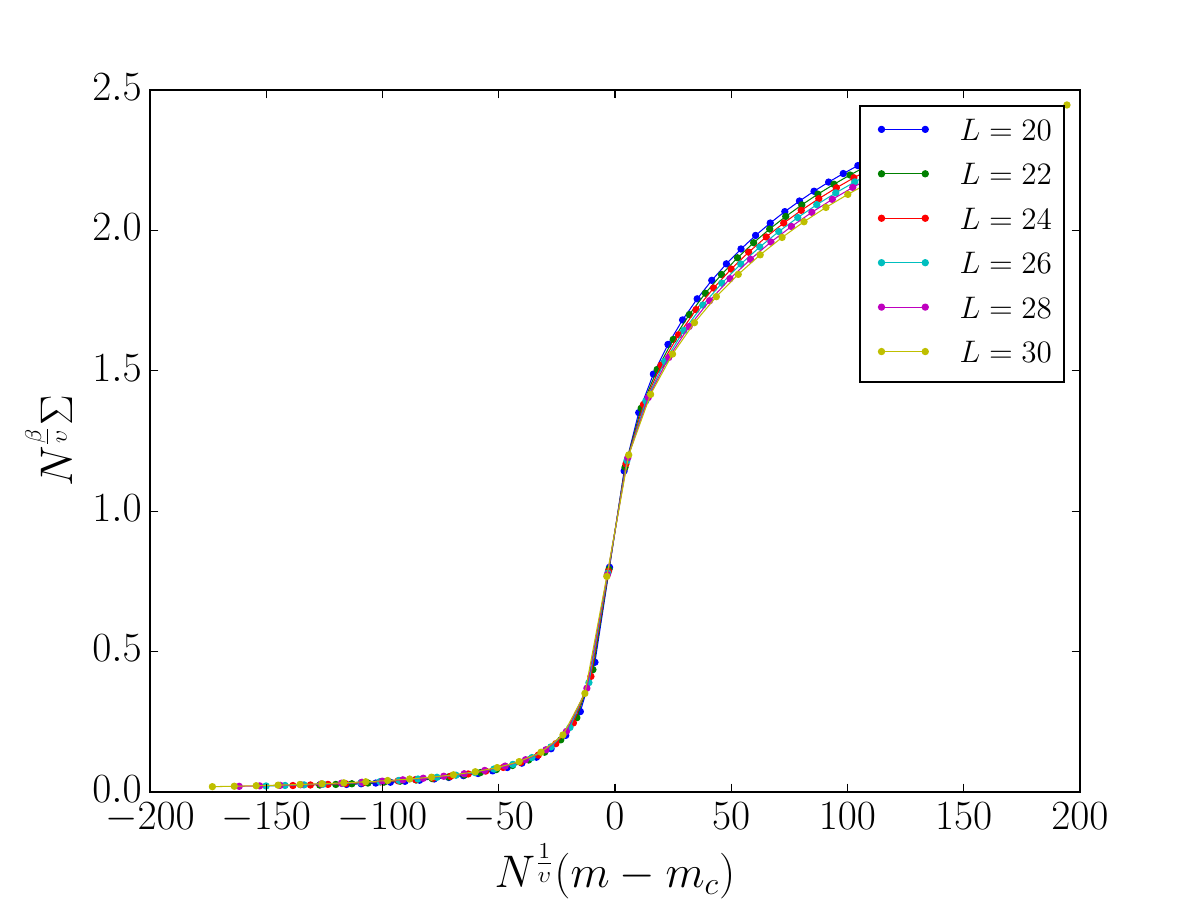}}
\caption{$\mathbb{Z}_2$-model. (a) Order parameter $\varSigma$ as function of $m$, for different $L$; (b) Scaling of $\lambda(x)$ close to the transition point.}
\end{figure}

For $t=0$ the energy per pair of these two states can be calculated exactly:
\begin{equation}
\frac{E_{\mathrm{polarized}}}{L}=m+2 \left(\frac{\pi}{4}\right),
\end{equation} 
while
\begin{equation}
\frac{E_{\mathrm{alternating}}}{L}=-m+2 \left(\frac{\pi}{2}\right).
\end{equation} 
Thus a first order phase transitions occurs at $m_0^{(2)}=0$. For $t\neq0$ we look for
the phase transition by numerically calculating the observable $\varSigma$ as function of $m$ and performing a finite-size scaling of the universal function that describes the order parameter. 
Fig.~\ref{fig:scaling_par_ordine1_z2} displays $\varSigma$ for $t=2\pi/2=\pi$, from which we can calculate the critical value $m_{c}=0.016\pm0.025$, while Fig.~\ref{fig:scaling_par_ordine2_z2} shows the corresponding universal function. 

As for the odd-$n$ case, we can numerically evaluate the critical value of the mass for different values of $t$ and get the fit of the function $m_c(t)$ according to Eq.~(\ref{formfit})
\begin{align}
&m_{0}^{(2)}=0.004\pm0.001, \\
 &\alpha_2=(8\pm5)\cdot10^{-6},\\
&\beta_2=0.0149\pm0.0003 .
\end{align} 
From these values, we can see that \emph{both} coefficients are very small, the dominant one being associated with the linear term. This is indeed one issue of the $\mathbb{Z}_2$-model that, as we will presently see, is not shared by higher $n$-models. 

Taking into account this peculiarity and considering that we want to perform a large-$n$ limit by using at least three different values of $n$ both in the even and odd case, we have performed a similar analysis for the $\mathbb{Z}_4$, $\mathbb{Z}_6$ and $\mathbb{Z}_8$-models, whose results are summarized in the Appendix. By fitting these data, we obtain the coefficients $m_0^{(n)}$, 
$\alpha_n$ and $\beta_n$ of Eq.~(\ref{formfit}) as given in Table~\ref{table:msum}. As for the odd case, we find an excellent agreement of the numerical value for $m_0^{(n)}$ with the theoretically predicted value, $m_0=0$, and confirm that the dominant term is the one containing $\sqrt{t}$, as expected from Eq.~(\ref{mt0}). 

\section{Large-$n$ limit}
\label{sec:compara}

\begin{table}
\centering{}
\begin{tabular}{|c|c|c|c|}
\hline 
$n$ & $m_{0}^{(n)}$ & $\alpha_n$ & $\beta_{n}$ \tabularnewline
\hline 
\hline 
2 & $0.004\pm0.001$& $(8\pm5)\cdot10^{-6}$ & $0.0149\pm0.0003$\tabularnewline
\hline 
3 & $-1.0472\pm0.0001$&$-0.603\pm0.001$ & $-0.02\pm0.01$\tabularnewline
\hline 
4& $(-3\pm1)\cdot10^{-7}$&$0.626\pm0.005$ & $0.0290\pm0.0006$\tabularnewline
\hline 
5 & $-0.628\pm0.001$&$-0.494\pm0.004$ & $-0.015\pm0.001$\tabularnewline
\hline 
6 & $(-7.2\pm0.1)\cdot10^{-6}$& $0.543\pm0.005$ & $0.026\pm0.001$\tabularnewline
\hline 
7 &$ -0.448\pm0.001$ & $ -0.435\pm0.003$ & $0.004 \pm0.001$\tabularnewline
\hline 
8 & $(1.8\pm0.1)\cdot10^{-7}$ &$0.503\pm0.004$& $0.022 \pm0.001$\tabularnewline
\hline 
\end{tabular}\caption{Parameters of the numerical fit of the critical mass as a function of $t$, according to the formula $m_c(t) = m_0 + \alpha\sqrt{t} + \beta \, t$, for the various $\mathbb{Z}_{n}$-models with $n=2\div 8$. } 
\label{table:msum} 
\end{table}

\begin{figure}
\centering
\includegraphics[width=0.5\textwidth]{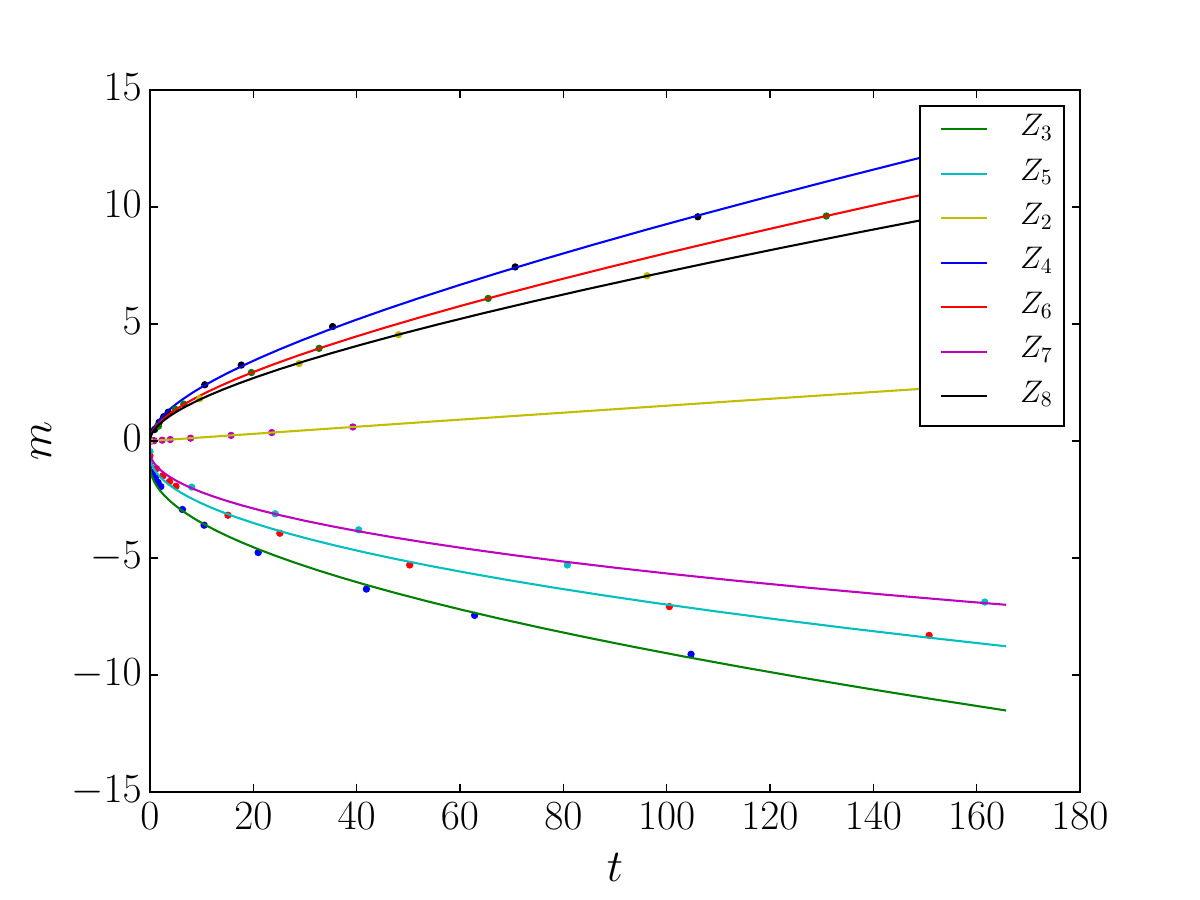}
\caption{Plot of $m_c(t)$ for the various $\mathbb{Z}_{n}$-models. The points are the numerical data of Table~\ref{table:msum} while  the fits (continuous lines) yield $m_c(t)= m_0^{(n)}+ \alpha_n \sqrt{t}$.}
\label{fig:comparazione_diagrammi} 
\end{figure}

\begin{figure}
\centering
\subfigure[\label{fig:alphaodd}]{\includegraphics[width=0.45\textwidth]{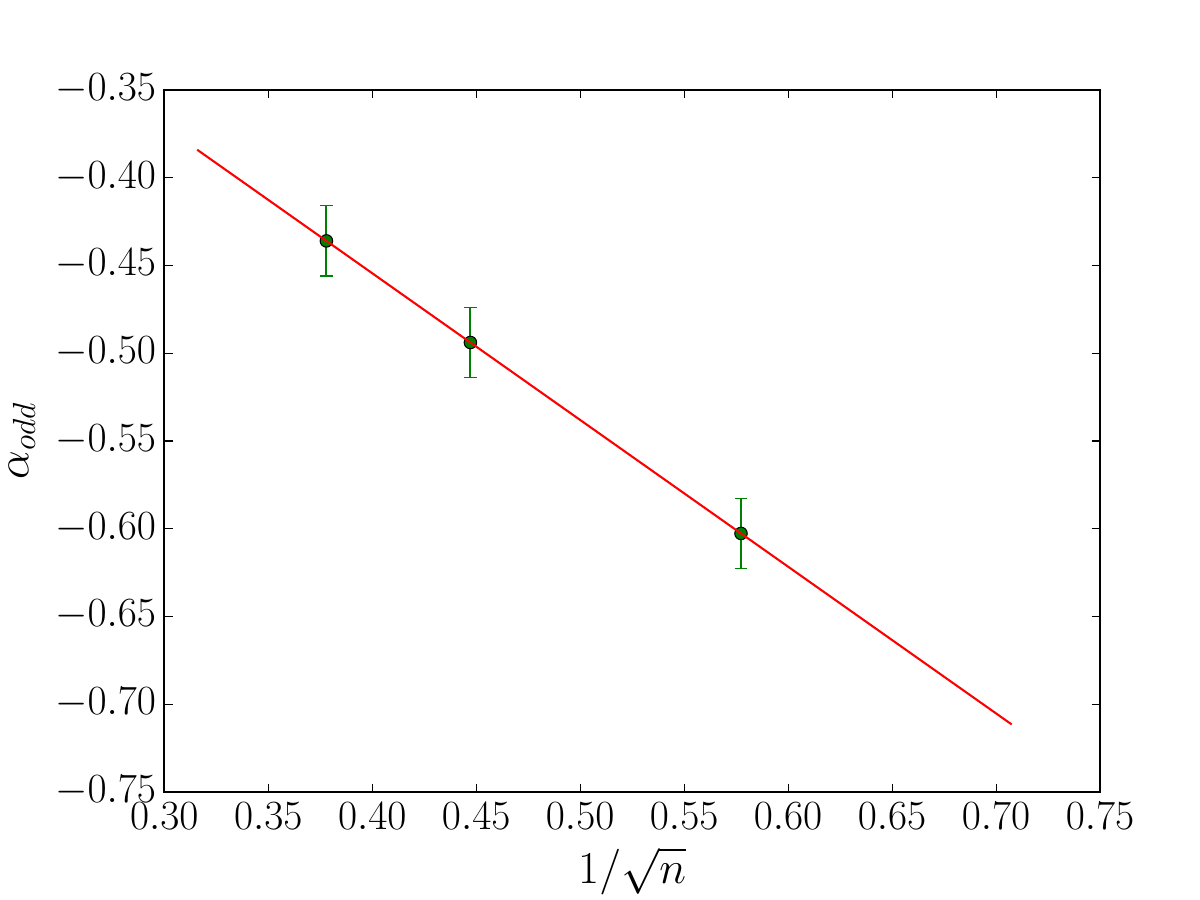}}
\subfigure[\label{fig::alphaeven}]{\includegraphics[scale=0.4]{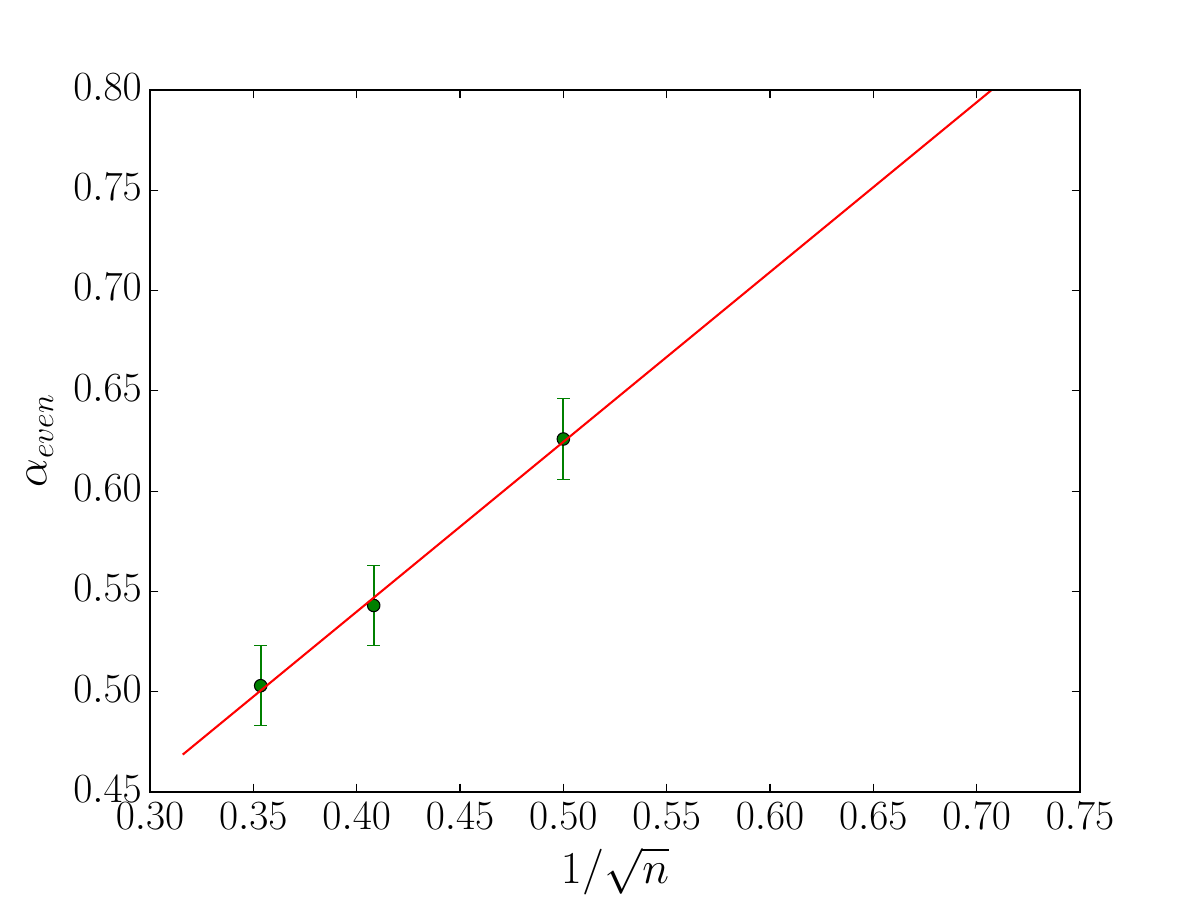}}
\caption{Scaling with $1/\sqrt{n}$ of the coefficient $\alpha_n$, for (a) $n=3,5,7$ and (b) $n=4,6,8$.} \label{fig:alfa}
\end{figure}

The values of $m_c(t)$  for $n=2 \div 8$, are summarized in Table~\ref{table:msum}  and plotted in Fig.~\ref{fig:comparazione_diagrammi}. Leaving aside the peculiar $n=2$ case, these data clearly show that  the critical mass actually shows a square-root dependence on $t$
\begin{equation}
\label{mct}
m_c(t) = m_0^{(n)} + \alpha_n \sqrt{t},
\end{equation} 
where, for any $n$, the critical mass at $t=0$ can be calculated analytically (with $g^2 a=2$) according to the formula 
\begin{equation}
m_{0}^{(n)}=\begin{cases}
- \frac{\pi}{n}& n \; \textrm{odd} \\
0 & n \; \textrm{even}
\end{cases} , 
\end{equation}
and vanishes in the large-$n$ limit.
The coefficients $\alpha_n$ can be read from the third column of Table~\ref{table:msum}. As we can see from Fig.~\ref{fig:alfa}, they obey the scaling 
\begin{align}
& \alpha_n \simeq b + d / \sqrt{n} , \label{d83a}\\
\label{d83}
&d=\begin{cases} -0.83\pm0.10 & \qquad n \; \textrm{odd} \\
+0.84\pm0.17 & \qquad n \; \textrm{even}
\end{cases}
\end{align}
and $b=0$ within numerical error in both cases.
Except for the different sign, which is due to the fact that the even $n$-models do not admit a zero electric field, these two values are the same. 

Thus, combining~(\ref{mtn}) and~(\ref{d83a}), we conclude that the continuous U$(1)$ theory exhibits a phase transition at the critical mass ($t=1$)
\begin{equation}
m_c= \alpha = \lim_{n\rightarrow \infty} \alpha_n \sqrt{\frac{n}{2\pi} }=  \frac{d}{\sqrt{2\pi} } \simeq \pm 0.33,
\end{equation}
with the sign depending on the charge sector. 
This value is in very good agreement with the estimates $m_{c}/g=0.33(2)$, obtained by using a lattice Hamiltonian approach~\cite{HKCM}, and $m_{c}/g=0.3335(2)$, obtained by studying the truncated $\mathbb Z$-model (at most at the first five loop levels)~\cite{byrnes2002}.

\section{Cold-atom simulator}
\label{sec:coldatom}

\begin{figure}
\centering
\includegraphics[width=0.48\textwidth]{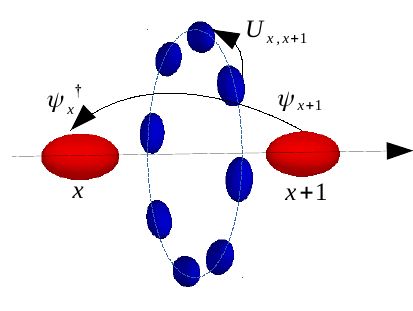}
\caption{Scheme of the physical system on which the $\mathbb{Z}_n$ models can be implemented. Between each pair of neighboring fermion sites (red spots), a single particle is bound to hop on a circlular lattice. In order to reproduce the gauge Hamiltonian~(\ref{eq:Schwinger}), hopping of fermions and of the particle on the link must be correlated.}\label{fig:lattice3d}
\end{figure}

The implementation of the $\mathbb{Z}_n$ models is complicated by the presence of the correlated hopping terms, related to elementary processes in which the hopping of a fermion to a nearest-neighboring site is always associated to an action on the link between the sites, which amounts at increasing the electric field in the case of hopping to the left, and decreasing it if the fermion hops to the right. The accuracy of correlated hopping terms is vital for any cold-atomic simulator of the described theories, since it guarantees that, once the system starts in the physical subspace, in which Gauss' law is satisfied, it will not leave this subspace during the evolution. However, in a quantum simulation, in which matter and gauge fields are encoded in the external and (possibly) internal degrees of freedom of cold atoms, Gauss' law does not emerge as a natural property. Implementation of correlated hopping and enforcement of Gauss' law are therefore still open problems. We can identify two possible ways to simulate the gauge variables with cold atoms:
\begin{itemize}
\item Gauge variables can be encoded in the internal degrees of freedom of single atoms trapped at intermediate positions between each couple of adjacent sites. Hopping of a fermion induces transition towards different states according to the hopping direction. This realization require a fine tuning of atomic transitions, as care must be taken in ensuring that all the allowed transition amplitudes between states with given fermion occupation numbers and electric field are equal. Moreover, the pure-gauge term requires that the energy levels of the intermediate atoms at $a\to\infty$ are quadratically spaced.
\item Gauge variables can be encoded in an external, transverse degree of freedom. A possible interesting implementation arises from the possibility of trapping cold atoms in circular lattices, obtained by interaction with Laguerre-Gauss laser modes~\cite{aoc}. The scheme is represented in Figure \ref{fig:lattice3d}, where the red spots represent the bottoms of potential wells in which the fermions are trapped, while the blue ones host one particle per link (statistics is immaterial), which can hop through neighboring sites of each circle (identified with eigenstates of the electric field), but cannot hop towards other links due to a large energy barrier. The equal amplitude of hopping between sites on the circle arises from a natural circular symmetry, and the pure-gauge term can be implemented by adding an external potential that properly varies along one of the transverse directions. 
\end{itemize}
In both cases, the Gauss law could be implemented either by tailoring the transition amplitudes in order to enhance correlated hopping and suppress the forbidden terms, or by adding an energy or noise penalty to the states that violate Gauss' law~\cite{simul2,simul11,zohar2011}. In the latter case, the desired interaction Hamiltonian can emerge as a higher-order effective dynamics~\cite{NEFMPP}.

\section{Conclusions}
\label{sec:concl}

We have investigated  discrete $\mathbb{Z}_n$  models, that approximate QED in one dimension (Schwinger model). In these models the electric field can take a finite number of values, and one important common feature is the preservation of the unitarity of the comparator. Thus, we have put the large-$n$ limit on a firm mathematical ground, adding novel rigorous results in the field of quantum simulations of gauge fields, that may soon find experimental verifications in cold-atomic systems.

In particular, we have unveiled the presence of phase transitions, whose features depend in an interesting way on whether $n$ is even or odd. Although the details of these transitions depend on $n$, their universality class, as well as some of their main features, are $n$-independent, so that by looking at the large $n$ limit, in which $\mathbb{Z}_n \to U(1)$, one can establish the presence of a phase transition for one-dimensional lattice QED, and extract crucial information.

A possible implementation of $\mathbb{Z}_n$ models on a cold-atom simulator, discussed in~\cite{NEFMPP} and reviewed in Section \ref{sec:coldatom}, relies on the identification of the discrete values taken by the electric field with some suitable additional degrees of freedom of the simulator~\cite{aoc}. A realization appears realistic and would be important to elucidate some important features of one-dimensional QED. Clearly, $1+1$-dimensional models have to be considered as toy-models with respect to the more realistic $3+1$-dimensional ones, but the possibility of using quantum simulators for the investigation of collective and non-perturbative features of gauge theories would enable us to shed new light on old problems, and provide new insights on crucial but still unsolved questions.

The dynamics of lattice U$(1)$ gauge theories has been recently experimentally investigated in a few-qubit trapped-ion quantum computer~\cite{martinez2016}. Therefore, among the computational perspectives, it is worth mentioning the study of real time dynamical phenomena, such as string breaking~\cite{marcos2013,montangero2015,buyens_prx}, and the time evolution of localized excitations in the different phases identified in this work.

\section*{Acknowledgments} The authors are grateful to Fabio Ortolani for his precious help with the DMRG code.
EE, PF, GM and SP are partially supported by INFN through the project ``QUANTUM''.
FVP is supported by INFN through the project ``PICS''.
PF is partially supported by the Italian National Group of Mathematical Physics (GNFM-INdAM).

\appendix
\section{Additional information on the phase transition}
\label{addinfo}

\begin{figure}[h]
\centering
\includegraphics[width=0.5\textwidth]{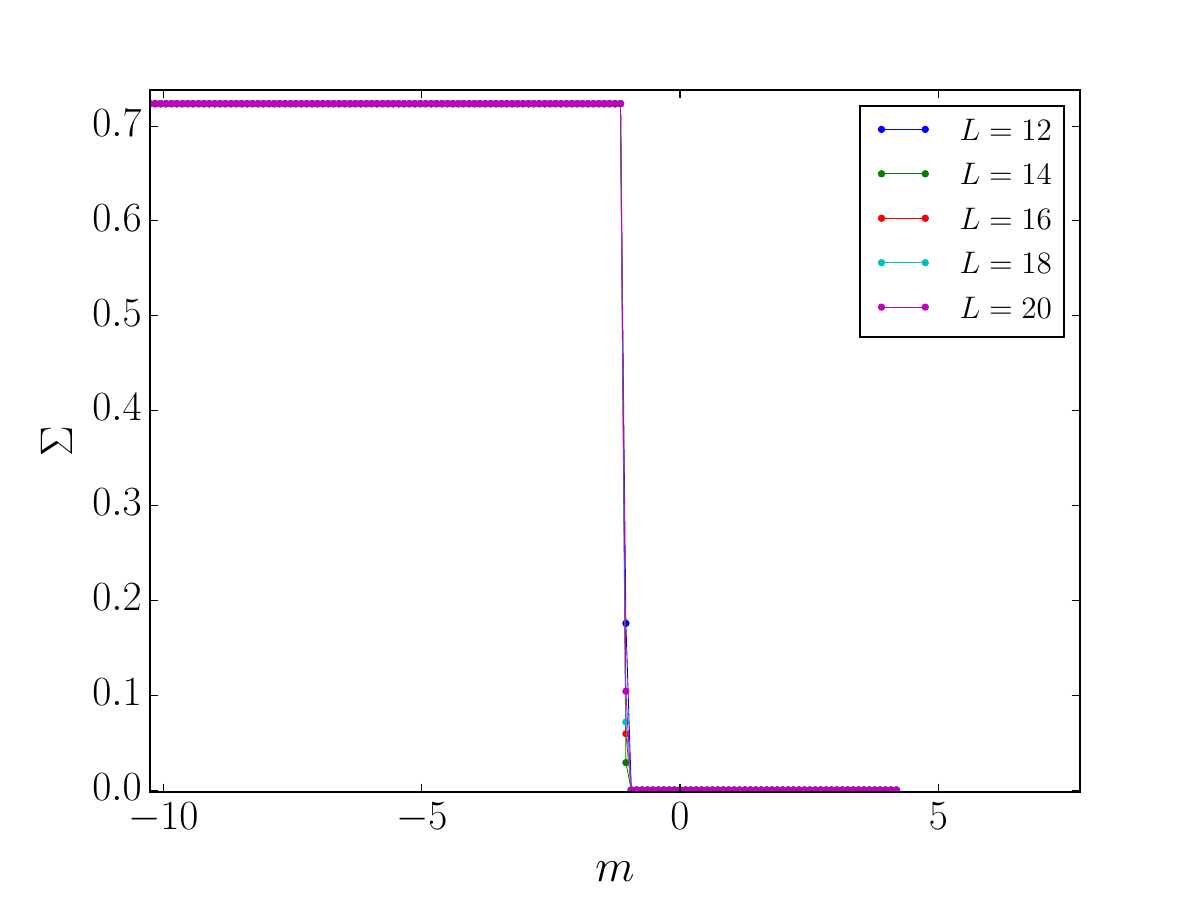}
\caption{$\mathbb{Z}_3$-model. $\varSigma$ as function of $m$ at $t=0$, for different system size $L$.}
\label{fig:par_ordine_t_0.0}
\end{figure}

We give here additional details on the phase transition, as well as numerical figures, for different $\mathbb{Z}_n$-models. \\
Let us start with $n=3$. 
For $t=0$, the transition is sharp for every system size, as can be seen in Fig.~\ref{fig:par_ordine_t_0.0}. As explained at the end of Sec.\ \ref{sec:zbb}, the system undergoes here a first order phase transition between the Dirac sea and the mesonic state shown in Figs.\  \ref{fig:vacuum} and \ref{fig:mesons}. Our numerical findings for $m_c(t)$ as a function of $t$ are summarized in Table~\ref{table:mt3}.  \\
Our numerical findings for $m_c(t)$ for the $\mathbb{Z}_5$ and $\mathbb{Z}_7$-models are given in 
Tables~\ref{table:mt5} and \ref{table:mt7}, respectively. 
Those for the $\mathbb{Z}_2$, $\mathbb{Z}_4$, $\mathbb{Z}_6$ and $\mathbb{Z}_8$-models are reported in Tables~\ref{table:mt2}, \ref{table:mt4}, \ref{table:mt6}, and \ref{table:mt8}, respectively.\\
All the values given in the Tables are plotted in Fig.\ \ref{fig:comparazione_diagrammi}, to yield the fit in Eq.~(\ref{mct}).

\begin{table*}
\begin{tabular}{|c|c|c|c|c|c|c|c|c|c|c|c|}
\hline
$t$ & $0.000$ & $0.523$ &$1.047$ & $1.571$ &$2.094$ &$6.283$ & $10.472$ &$20.944$ & $41.888$ & $62.832$ & $104.720$ 
\tabularnewline
\hline
$m_{c}$ & $-1.047$ & $-1.340$ & $-1.571$ & $-1.770$ & $-1.948$ & $-2.927$ & $-3.596$ & $-4.767$ & $-6.329$ & $-7.449$ & $-9.115$
\tabularnewline
\hline 
\end{tabular} \caption{$\mathbb{Z}_{3}$-model. Critical values  $m_{c}(t)$ for different values of $t$. The error is always $0.025$.}\label{table:mt3} 
\vskip0.5cm
\begin{tabular}{|c|c|c|c|c|c|c|c|c|c|c|c|}
\hline
$t$ & $0.000$ & $1.257$ & $2.513$ & $3.769$ & $5.026$ & $15.080$ & $25.133$ & $50.265$ & $100.531$ & $150.796$ 
\tabularnewline
\hline
$m_{c}$ & $-0.628$ & $-1.181$ & $-1.483$ & $-1.734$ & $-1.935$ & $-3.171$ & $-3.941$ & $-5.298$ & $-7.077$ & $-8.309$
\tabularnewline
\hline 
\end{tabular} \caption{$\mathbb{Z}_{5}$-model. Critical values $m_{c}(t)$ for different values of $t$.The error is always $0.025$.}\label{table:mt5} 
\vskip0.5cm
\begin{tabular}{|c|c|c|c|c|c|c|c|c|c|c|c|}
\hline
$t$ & $0.000$ & $8.078$ & $24.235$ & $40.392$ & $80.784$ & $161.568$ 
\tabularnewline
\hline
$m_{c}$ & $-0.448$ & $-1.971$ & $-3.110$  & $-3.797$ & $-5.299$ & $-6.883$
\tabularnewline
\hline 
\end{tabular} \caption{$\mathbb{Z}_{7}$-model. Critical values $m_{c}(t)$ for different values of $t$. The error is always $0.025$.}\label{table:mt7} 
\vskip0.5cm
\begin{tabular}{|c|c|c|c|c|c|c|c|c|c|c|c|}
\hline
$t$ & $0.000$ & $0.196$ & $0.392$ & $0.589$ & $0.785$ & $2.356$ & $3.927$ & $7.854$ & $15.708$ & $23.562$ & $39.269$ 
\tabularnewline
\hline
$m_{c}$ & $0.000$ & $0.012$ & $0.013$ & $0.014$ & $0.016$ & $0.035$ & $0.062$ & $0.122$ & $0.239$ & $0.361$ & $0.601$
\tabularnewline
\hline 
\end{tabular} \caption{$\mathbb{Z}_{2}$-model. Critical values $m_{c}(t)$ for different values of $t$. The error is always $0.025$. }\label{table:mt2} 
\vskip0.5cm
\begin{tabular}{|c|c|c|c|c|c|c|c|c|c|c|c|}
\hline
$t$ & $0.000$ & $0.884$ & $1.767$ & $2.651$ & $3.534$ & $10.603$ & $17.671$ & $35.342$ & $70.685$ & $106.029$
\tabularnewline
\hline
$m_{c}$ & $0.000$ & $0.491$ & $0.795$ & $1.039$ & $1.233$ & $2.403$ & $3.244$ & $4.887$ & $7.439$ & $9.581$
\tabularnewline
\hline 
\end{tabular} \caption{$\mathbb{Z}_{4}$-model. Critical values $m_{c}(t)$ for different values of $t$. The error is always $0.025$..}\label{table:mt4} 
\vskip0.5cm
\begin{tabular}{|c|c|c|c|c|c|c|c|c|c|c|c|}
\hline
$t$ & $0.000$ & $1.636$ & $4.909$ & $6.545$ & $19.635$ & $32.725$ & $65.449$ & $130.889$ 
\tabularnewline
\hline
$m_{c}$ & $0.000$ & $0.635$ & $1.355$ & $1.577$ & $2.925$ & $3.959$ & $6.093$ & $9.614$
\tabularnewline
\hline 
\end{tabular} \caption{$\mathbb{Z}_{6}$-model. Critical values $m_{c}(t)$ for different values of $t$. The error is always $0.025$.}\label{table:mt6} 
\vskip0.5cm
\begin{tabular}{|c|c|c|c|c|c|c|c|c|c|c|c|}
\hline
$t$ & $0.000$ & $9.621$ & $28.863$ & $48.106$ & $96.211$ 
\tabularnewline
\hline
$m_{c}$ & $0.000$ & $1.809$ & $3.309$ & $4.541$ & $7.062$
\tabularnewline
\hline 
\end{tabular} \caption{$\mathbb{Z}_{8}$-model. Critical values $m_{c}(t)$ for different values of $t$. The error is always $0.025$.}\label{table:mt8} 
\end{table*}


\begin{thebibliography}{10}


\bibitem{rothe1992lattice}
H. J. Rothe, \textit{Lattice gauge theories} (World Scientific, Singapore, 1992).

\bibitem{montvay1997quantum}
I. Montvay and G. M\"unster, {\it Quantum Fields on a Lattice} (Cambridge University Press, Cambridge, 1994).

\bibitem{wilsonlgt}
K. Wilson, Phys. Rev. D {\bf 10}, 2445 (1974).

\bibitem{kogut1975hamiltonian}
J. B. Kogut and L. Susskind, Phys. Rev. D {\bf 11}, 395 (1975).

\bibitem{susskind1977lattice}
L. Susskind, Phys. Rev. D {\bf 16}, 3031 (1977).

\bibitem{kogut1979introduction}
J. B. Kogut, Rev. Mod. Phys. {\bf 51}, 659 (1979).


\bibitem{bdz}
I. Bloch, J. Dalibard, and W. Zwerger, Rev. Mod. Phys. {\bf 80}, 885 (2008).

\bibitem{qsim1}
M. Lewenstein, A. Sanpera and V. Ahufinger, \textit{Ultracold
Atoms in Optical Lattices: Simulating Quantum Many-Body Systems}
(Oxford University Press, New York, 2012).

\bibitem{qsim2}
J. I. Cirac and P. Zoller, Nat. Phys. {\bf 8}, 264 (2012).

\bibitem{qsim3}
I. Bloch, J. Dalibard, and S. Nascimb\`ene, Nat. Phys {\bf 8}, 267 (2012).

\bibitem{qsim4}
R. Blatt and C. F. Roos, Nat. Phys. {\bf 8}, 277 (2012).



\bibitem{simul1}
E. Kapit and E. Mueller, Phys. Rev. A {\bf 83}, 033625 (2011).

\bibitem{simul4}
E. Zohar, J. I. Cirac, and B. Reznik, Phys. Rev. Lett. {\bf 109}, 125302 (2012).

\bibitem{simul3}
L. Tagliacozzo, A. Celi, P. Orland, and M. Lewenstein, Nat. Commun. {\bf 4}, 2615 (2013).

\bibitem{simul5}
K. Kasamatsu, I. Ichinose, and T. Matsui, Phys. Rev. Lett. {\bf 111}, 115303 (2013).

\bibitem{simul6}
D. Banerjee, M. B\"ogli, M. Dalmonte, E. Rico, P. Stebler, U. J. Wiese, and P. Zoller, Phys. Rev. Lett. {\bf 110}, 125303 (2013).

\bibitem{simul7} 
L. Tagliacozzo, A. Celi, A. Zamora, and M. Lewenstein, Ann. Phys. (Amsterdam) {\bf 330}, 160 (2013).

\bibitem{simul9}
E. Zohar, J. I. Cirac, and B. Reznik, Phys. Rev. A {\bf 88}, 023617 (2013).

\bibitem{simul11}
K. Stannigel, P. Hauke, D. Marcos, M. Hafezi, S. Diehl, M. Dalmonte, and P. Zoller, Phys. Rev. Lett. {\bf 112}, 120406 (2014).

\bibitem{zoharreview}
E. Zohar, J. I. Cirac, and B. Reznik, \textit{Quantum Simulations of Lattice Gauge Theories using Ultracold Atoms in Optical Lattices}, Rep. Prog. Phys. {\bf 79}, 014401 (2016).



\bibitem{simul10} 
P. Hauke, D. Marcos, M. Dalmonte, and P. Zoller, Phys. Rev. X {\bf 3}, 041018 (2013).

\bibitem{KCB} 
S. K\"uhn, J. I. Cirac, and M.C. Ba\~nuls, Phys. Rev. A \textbf{90}, 042305 (2014).

\bibitem{NEFMPP}
S. Notarnicola, E. Ercolessi, P. Facchi, G. Marmo, S. Pascazio and F. V. Pepe, J. Phys. A: Math. Theor. \textbf{48}, 30FT01 (2015).

\bibitem{simul12}  
V. Kasper, F. Hebenstreit, F. Jendrzejewski, M K Oberthaler, and J. Berges, New J. Phys. {\bf 19} (2017), 023030.


\bibitem{cmr} 
A. Celi, P. Massignan, J. Ruseckas, N. Goldman, I. B. Spielman, G. Juzeli\={u}nas, and M. Lewenstein, Phys. Rev. Lett. \textbf{112}, 043001 (2014).

\bibitem{ytterbium} 
G. Pagano, M. Mancini, G. Cappellini, P. Lombardi, F. Sch\"afer, H. Hu, X.-J. Liu, J. Catani, C. Sias, M. Inguscio, and L. Fallani, Nat. Phys. \textbf{10}, 198 (2014).

\bibitem{mpi_sun} 
F. Scazza,	C. Hofrichter,	M. H\"ofer,	P. C. De Groot,	I. Bloch, and S. F\"olling, Nat. Phys \textbf{10}, 779 (2014).

\bibitem{fallani} 
M. Mancini, G. Pagano, G. Cappellini, L. Livi, M. Rider, J. Catani, C. Sias, P. Zoller, M. Inguscio, M. Dalmonte, L. Fallani, Science \textbf{349}, 1510 (2015).

\bibitem{lcd} 
L. F. Livi, G. Cappellini, M. Diem, L. Franchi, C. Clivati, M. Frittelli, F. Levi, D. Calonico, J. Catani, M. Inguscio, L. Fallani, Phys. Rev. Lett. \textbf{117}, 220401 (2016).

\bibitem{martinez2016} 
E.A. Martinez, C.A. Muschik, P. Schindler, D. Nigg, A. Erhard, M. Heyl, P. Hauke, M. Dalmonte, T. Monz, P. Zoller, and R. Blatt, Nature {\bf 534}, 516 (2016).

\bibitem{DMRG1}
U. Schollw\"{o}ck, Rev. Mod. Phys. \textbf{77}, 259 (2005).

\bibitem{MPS}
R. Orus, Annals of Physics \textbf{349}, 117 (2014).


\bibitem{simul2} 
D. Banerjee, M. Dalmonte, M. M{\"u}ller, E. Rico, P. Stebler, U. J. Wiese, and P. Zoller, Phys. Rev. Lett. {\bf 109}, 175302 (2012).

\bibitem{BCJC} 
M.C. Ba\~nuls, K. Cichy, K. Jansen, J.I. Cirac, JHEP \textbf{11}, 158 (2013).

\bibitem{rico2016}  
E. Rico, T. Pichler, M. Dalmonte, P. Zoller, and S. Montangero, Phys. Rev. Lett. {\bf 112}, 201601 (2014).

\bibitem{montangero2015} 
T. Pichler, M. Dalmonte, E. Rico, P. Zoller, and S. Montangero, Phys. Rev. X {\bf 6}, 011023 (2016).

\bibitem{schwinger_mps} 
B. Buyens, J. Haegeman, F. Hebenstreit, F. Verstraete and K. Van Acoleyen, \textit{Real-time simulation of the Schwinger effect with Matrix Product States}, arXiv:1612.00739 (2016).

\bibitem{buyens_prx} 
B. Buyens, J. Haegeman, H. Verschelde, F. Verstraete, K. Van Acoleyen, Phys. Rev. X \textbf{6}, 041040 (2016).

\bibitem{buyens} 
B. Buyens, S. Montangero, J. Haegeman, F. Verstraete and K. Van Acoleyen, 
Phys. Rev. D \textbf{95}, 094509 (2017).

\bibitem{qlm1}
D. Horn, Phys. Lett. {\bf 100B}, 149 (1981).

\bibitem{qlm2}
P. Orland and D. Rohrlich, Nucl. Phys. {\bf B338}, 647 (1990).

\bibitem{qlm3}
S. Chandrasekharan and U. J. Wiese, Nucl. Phys. {\bf B492}, 455 (1997).

\bibitem{qlm4wiese} 
U. J. Wiese, Annalen der Physik {\bf 525}, 777 (2013).




\bibitem{aoc} 
L. Amico, A. Osterloh, and F. Cataliotti, Phys. Rev. Lett. {\bf 95}, 063201 (2005).

\bibitem{schwinger2001quantum}
J. Schwinger and B. G. Englert, \textit{Quantum mechanics: symbolism
of atomic measurements} (Springer, Berlin, 2001).

\bibitem{melnikov}
K. Melnikov and M. Weinstein, Phys. Rev. D \textbf{62}, 094504 (2000).

\bibitem{weyl1950theory}
H. Weyl \textit{The theory of groups and quantum mechanics}
(Courier Dover Publications, 1950).


\bibitem{coleman2} 
S. Coleman, Ann. Phys. \textbf{101}, 239 (1976).

{\textcolor{red}{
\bibitem{manton} N. S. Manton, Ann. Phys. \textbf{159}, 220 (1985).
\bibitem{wipf} C. Kiefer and A. Wipf, Ann. Phys. \textbf{236}, 241 (1994).
}}

\bibitem{DMRG2}
C. Degli Esposti Boschi, F. Ortolani, Eur. Phys. J. B \textbf{41}, 503 (2004).

\bibitem{Cardy}
J. Cardy, \textit{Scaling and renormalization in statistical physics}
(Cambridge University Press, 1996).


\bibitem{shimizu_kuramashi}
Y. Shimizu and Y. Kuramashi, Phys. Rev. D \textbf{90}, 014508 (2014); Y. Shimizu and Y. Kuramashi, Phys. Rev. D \textbf{90}, 074503 (2014).

\bibitem{byrnes2002} T. M. R. Byrnes, P. Sriganesh, R. J. Bursill, and C. J. Hamer, Phys. Rev. D {\bf 66}, 013002 (2002); Nucl. Physics B Proceedings Supplements {\bf 109}, 202 (2002).

\bibitem{sachs_wipf}
I. Sachs, A. Wipf, Helv. Phys. Acta \textbf{65}, 652 (1992)

\bibitem{cardy_calabrese}
P. Calabrese and J. Cardy, J. Stat. Mech. \textbf{0406}, 002 (2004). 

\bibitem{henkel}
M. Henkel, \textit{Conformal Invariance and Critical Phenomena} (Springer, New York, 1999).

\bibitem{HKCM}
C. J. Hamer, J. Kogut, D. P. Crewther and M. M. Mazzolini, Nucl. Phys. B \textbf{208}, 413 (1982).

\bibitem{marcos2013} D. Marcos, P. Rabl, E. Rico, and P. Zoller, Phys. Rev. Lett. 111, 110504 (2013).

\bibitem{zohar2011}
E. Zohar and B. Reznik, Phys. Rev. Lett. \textbf{107}, 275301 (2011).


\end{thebibliography}
\end{document}